\documentclass[aps,prd,preprintnumbers,nofootinbib,floatfix,epsfig]{revtex4-1}
\usepackage{dcolumn}
\usepackage{subfigure}
\usepackage{graphicx}
\usepackage{color}
\usepackage{amsmath}
\usepackage{amssymb}
\usepackage{amsthm}
\usepackage{latexsym}
\usepackage[hmargin=1.8cm,vmargin=1.8cm]{geometry}
\usepackage{bm}
\usepackage{multirow}
\usepackage{cancel}
\usepackage{hyperref}
\usepackage{tabularx}
\begin{document}
\vspace*{2cm}
\title{Multi-photon production in the Type-I 2HDM}

\author{Abdesslam Arhrib}
\email[]{aarhrib@gmail.com}
\affiliation{\small Facult\'e des Sciences et Techniques, Abdelmalek Essaadi University, B.P. 416, Tangier, Morocco}

\author{Rachid Benbrik}
\email[]{r.benbrik@uca.ac.ma}
\affiliation{\small LPHEA, Semlalia, Cadi Ayyad University, Marrakech, Morocco}
\affiliation{\small MSISM Team, Facult\'e Polydisciplinaire de Safi,
Sidi Bouzid, BP 4162,  Safi, Morocco.}
\author{Stefano Moretti}
\email[]{s.moretti@soton.ac.uk}
\affiliation{\small School of Physics and Astronomy, University of Southampton,\\
	Southampton, SO17 1BJ, United Kingdom}

\author{Abdessamad Rouchad}
\email[]{abdessamad.rouchad@edu.uca.ac.ma}
\affiliation{\small LPHEA, Semlalia, Cadi Ayyad University, Marrakech, Morocco}

\author{Qi-Shu Yan}
\email[]{yanqishu@ucas.ac.cn}
\affiliation{\small 
School of Physics Sciences, University of Chinese 
Academy of Sciences, Beijing 100039, P.R China}
\affiliation{\small 
Center for future high energy physics,  Chinese  
Academy of Sciences, Beijing 100039, P.R. China}

\author{ Xianhui Zhang}
\email[]{ xianhuizhang@ucas.ac.cn}
\affiliation{\small 
School of Physics Sciences, University of Chinese 
Academy of Sciences, Beijing 100039, P.R China}

\date{\today}

\vspace*{-3cm}

\begin{abstract}
	This paper presents a study of a possible contribution to a Higgs
        boson signal
	in the $hh\to \gamma\gamma\gamma\gamma$ channel due to $H\to hh$
        decays, in
	the framework of the CP-conserving 2-Higgs Doublet Model  Type-I
        (2HDM-I), 
	where the heavier of the two CP-even Higgs bosons defined herein, $H$,
        is the
	SM-like Higgs state observed with a mass of 125 GeV at the Large Hadron
	Collider (LHC). We perform a broad scan of the 2HDM-I parameter space,
        in	presence of 
	both up-to-date theoretical and experimental constraints, in order to
        extract
	the interesting regions yielding such a signal. Then, after validating
        our
	numerical framework against public 
	experimental analyses carried out at the LHC, we proceed to assess its
        scope
	in  constraining and/or extracting 
	the $gg\to H\to hh\to \gamma\gamma\gamma\gamma$ signal in presence of a
	sophisticated Monte Carlo (MC) simulation. We find that, over a
        substantial
	region of the 2HDM-I parameter space presently un-accessible, the LHC
        will
	be able to  establish such a potential signature in the next 2--3
        years.  
\end{abstract}

\maketitle

\section{Introduction} 
Following the discovery of a Higgs boson with a mass of 125 GeV \cite{atlas,cms}, henceforth
labeled by $H$, with characteristics similar to those of the predicted state
of the Standard Model (SM), experiments at the LHC have effectively begun to
probe Electro-Weak Symmetry Breaking (EWSB) dynamics.  The search channel that
mostly enabled discovery was the one involving $gg\to H$ production followed
by a  $H\to \gamma\gamma$ decay, thanks to its cleanliness in the hadronic
environment of the LHC and the sharp resolution in the di-photon invariant
mass achievable by the LHC detectors, despite this decay being actually very
subleading. Other Higgs signals were eventually established and studied in
detail in order to measure the $H$ state fundamental parameters, i.e., mass,
width and couplings, all broadly consistent with the SM picture. Furthermore,
comprehensive  analyses  investigating  the  spin  and parity  of  the
discovered particle  have finally  confirmed  its  most likely spin-0 and
Charge/Parity (CP)-even nature, again, well in line with SM predictions.

The EWSB dynamics implemented within the SM is minimal in nature, allowing for the existence of only one Higgs boson. However, this needs not be the preferred realisation chosen by Nature. Just like the gauge and Yukawa sectors are not, 
i.e., there are multiple spin-1 and spin-1/2 states, there is a case for
conceiving the possibility of an extended spin-0 sector too. As the Higgs
boson so far discovered emerges from a doublet representation, a meaningful
approach to surpass the SM is exemplified by 2HDMs \cite{2hdms,hunters,lee}, wherein a second (complex)
Higgs doublet is added to the fundamental field representations of the SM. Upon EWSB, this yields five Higgs boson states as physical
objects: two neutral CP-even ones ($h$ and $H$ with, conventionally,
$m_{H}>m_h$), one neutral CP-odd one ($A$) and two charged ones ($H^\pm$) 
\cite{gunion}.

In the light of the established nature of the Higgs boson signals at the LHC,
as mentioned above, in terms of its mass, width, couplings, spin and CP state,
there exists therefore the possibility that in a 2HDM the observed SM-like
Higgs state can be either the $h$ \cite{carena,bernon} or $H$ 
\cite{ferreira,bernon2} one. An intriguing possibility is
that Nature made the second choice, i.e., the heavy 2HDM CP-even state, so
that a pair of the light ones could appear as its decay products, as the $Hhh$
vertex is indeed allowed by the most general 2HDM scalar potential and
underlying symmetries, which in fact coincide with those of the SM, except
(possibly) for an additional  $Z_2$ one introduced to prevent (large) Flavour
Changing Neutral Currents (FCNCs) \cite{Glashow:1976nt,Branco:2011iw} 
that may otherwise emerge in presence of a
2HDM (pseudo)scalar sector. Such a mass hierarchy, i.e., $m_H>2m_h$, can
easily be realised over the parameter space of one particular realisation of a
2HDM, so-called type-I (2HDM-I), see next section for details, 
which in fact allows
for $h$ masses down to even 20--30 GeV, well compatible with both theoretical
and experimental constraints. 


However, the requirement that one out of $h$ or $H$ has physical
properties consistent with the observed Higgs boson state puts rather 
stringent bounds on the 2HDM parameter
space. For example, it is well known that, in a 2HDM, there exists a
`decoupling limit', where $m_{H,A,H^\pm}\gg m_Z$, 
$\cos(\beta-\alpha)\approx 0$ \cite{gunion} and the couplings of the $h$ state
to the SM particles are identical to those of the SM Higgs
boson. Alternatively, a 2HDM also possesses an `alignment limit', in which
either one of $h$  \cite{carena,bernon} or $H$  \cite{ferreira,bernon2} can
mimic the SM Higgs boson. This is a welcome feature, as we will be working in
a configuration of the 2HDM-I \cite{type1} parameter space close to the alignment limit
realised through the $H$ state, however, we will specifically be concentrating
on those parameters which enable the $h$ state to be (nearly) fermiophobic,
so that the $h\to \gamma\gamma$ decay (mediated by $W^\pm$ and $H^\pm$ boson loops)) can be dominant.

Hence, all this opens up the possibility of a rather spectacular 2HDM-I
signal, in the form of the following production and (cascade) decay process,
$gg\to H\to hh\to\gamma\gamma\gamma\gamma$, indeed relying upon the
aforementioned characteristic of photonic signals in the LHC
detectors. Clearly, the presence of two Higgs bosons as intermediate states in
such a signature induces a phase space suppression (with respect to the production
of a single Higgs state), however, this can be well compensated by the fact
that the $H\to hh$ transition is resonant and, as mentioned, di-photon decays
of the $h$ state can be dominant in the 2HDM-I \cite{type1} when occurring near its
fermiophobic limit. Furthermore, the knowledge of the $H$ mass (125 GeV in
our scenario), combined with the ability of reconstructing in each event
photon pairs with similar masses, the former thus enabling one to enforce the
$m_{\gamma\gamma\gamma\gamma}\approx 125$ GeV requirement and the latter  the
$m_{\gamma\gamma}\approx m'_{\gamma\gamma}$ one, allows us to exploit two
powerful kinematic handles in suppressing the background, again, bearing in
mind  the high mass resolutions achievable in photon mass reconstructions.

In the present study, in essence, we explore the discovery potential of a
light scalar Higgs boson $h$ of mass less than $m_H/2$ at the LHC Run 2 (hence
with Center-of-Mass (CM) energy $\sqrt{s}=13$ TeV and standard luminosity
conditions), where -- as explained -- $H$ represents the SM-like Higgs
state \cite{ferreira,bernon2}. Chiefly, we consider the production of a light $h$ pair indirectly via the
decay
\begin{equation}
H\to hh,
\end{equation}
with the production of $H$ via the standard mechanisms, which are dominated by
gluon-gluon fusion. In this connection, it is to be noted that the total
Branching Ratio (BR) of the SM-like Higgs boson to undetected Beyond the SM
(BSM) decay modes (BR$_{\rm BSM}$) is restricted by current Higgs data 
and predicted to be  \cite{brbsm}
\begin{equation}
{\rm BR}_{\rm BSM}\leq 0.34~\text{at~95\%~Confidence~Level~(CL).}
\end{equation}
That is, the presence of non-SM decay modes of SM-like Higgs boson is not
completely ruled out, which acts as a further motivation of our study.

In carrying out the latter, we borrow from existing experimental results.
The ATLAS collaboration carried out searches for new phenomena in events with
at least three photons at a CM energy of 8 TeV and with an integrated
luminosity of $20.3$ fb$^{-1}$. From the non-observation of any excess,
limits are set at 95\% CL on the rate of the relevant signal events in terms
of cross section multiplied by a suitable  BR combination  \cite{beta}
\begin{equation}
\sigma_{\rm BSM}\times\beta^\prime\leq 10^{-3}\sigma_{\rm SM}, 
\end{equation}
where $\beta^\prime={\rm BR}(H\to AA)\times {\rm BR}(A\to\gamma\gamma)^2$,
$\sigma_{\rm BSM}$ is the Higgs production cross section in a possible BSM
scenario and $\sigma_{\rm SM}$ is the same, but for the SM Higgs. The above
constraint sets an upper limit on $\beta^\prime$ as
\begin{equation}
\beta^\prime\leq 10^{-3},
\end{equation}
provided the Higgs state $H$ in the context of new physics phenomena is the
SM-like Higgs boson of mass 125 GeV. In particular, we will validate a
numerical toolbox that we have created to carry out a Monte Carlo (MC) analysis against
results published therein for the case of $H\to AA$ decays and extrapolate
them to the case of $H\to hh$ ones, which constitute the dominant
four photon signal in our case.

{The plan of the paper is as follows. In Section 2, 
we introduce 2HDMs \cite{hunters,2hdms,lee}
in general and describe in particular our construct (the 2HDM-I \cite{type1}),
including dwelling on the theoretical (see \cite{th20,th21,th23,th24,th27,th25}) 
and experimental (see later on) constraints placed upon its
parameter space. (Herein, we also comment on the fermiophobic limit of the 2HDM-I and
its experimental status.) Section 3 is devoted to present our numerical results for 
the (inclusive) four photons cross section and to motivate the selection of our Benchmark Points (BPs).
Then we move on to describe the numerical tools we have used
and the MC analysis carried out, including illustrating our results for the exclusive cross section
 in Section
4. We then conclude in Section 5. Some technical details of our calculations are presented in Appendix A.
}

\section{The 2HDM-I and its fermiophobic limit}
\subsection*{The 2HDM scalar potential}
The most general 2HDM scalar potential which is $SU(2)_L\otimes U(1)_Y$
invariant with a softly broken $Z_2$ symmetry can be written as
\begin{align}
V(\phi_1,\phi_2)&=m^2_{11}\phi^+_1\phi_1+m^2_{22}\phi^+_2\phi_2-[m^2_{12}\phi^+_1\phi_2+h.c]\nonumber\\
&+\frac{1}{2}\lambda_1(\phi^+_1\phi_1)^2+\frac{1}{2}\lambda_2(\phi^+_2\phi_2)^2+\lambda_3(\phi^+_1\phi_1)(\phi_2^+\phi_2)\\
&+\lambda_4(\phi_1^+\phi_2)(\phi_2^+\phi_1)+[\frac{1}{2}\lambda_5(\phi_1^+\phi_2)^2+{\rm h.c.}],\nonumber
\end{align}
where $\phi_1$ and $\phi_2$ have weak hypercharge $Y=+1$ while $v_1$ and
$v_2$ are their respective Vacuum Expectation Values (VEVs). Through the
minimisation conditions of the potential, $m_{11}^2$ and $m_{22}^2$ can be
traded for $v_1$ and $v_2$ and the tree-level mass relations allow the quartic
couplings $\lambda_{1-5}$ to be substituted by the four physical Higgs boson
masses and the neutral sector mixing term $\sin(\beta-\alpha)$, where $\beta$
is defined through $\tan\beta=v_2/v_1$ and $\alpha$ is the mixing angle
between the CP-even interaction states. Thus, in total, the Higgs sector of
the 2HDM has 7 independent parameters, which include $\tan\beta$,
$\sin(\beta-\alpha)$ (or $\alpha$), $m_{12}^2$ and 
the four physical Higgs boson masses.

As explained in the Introduction, the 2HDM possesses two alignment limits: one
with $h$ SM-like \cite{carena,bernon} and an other one with $H$ SM-like \cite{ferreira,bernon2}.
In the present study, we are interested in the alignment limit where 
$H$ is the SM-like Higgs boson discovered at CERN, which implies that 
$\cos(\beta-\alpha)\approx 1$. Then, we take 
$m_h<m_H/2\approx 62.5$ GeV, so that the decay channel $H\to hh$ 
would always be open.

From the above scalar potential one can derive the following 
triple scalar couplings needed for our study:
\begin{eqnarray}
Hhh &=& -\frac{1}{2}\frac{g c_{\beta-\alpha}}{m_W s^2_{2\beta}}\bigg[
(2 m^2_{h} + m^2_{H}) s_{2\alpha} s_{2\beta} -2 (3 s_{2\alpha}-s_{2\beta})
m^2_{12}\bigg], \nonumber \\
HAA &=& -\frac{g}{2m_W s^2_{2\beta}}\bigg[
(2 m^2_{A} - m^2_{H}) s_{2\beta}^2 c_{\beta-\alpha} +2 m^2_{H} s_{2\beta} s_{\beta+\alpha}
-4 m_{12}^2 s_{\beta+\alpha}  \bigg], \nonumber \\
hH^\pm H^\mp &=& \frac{1}{2}\frac{g}{m_W s^2_{2\beta}} \bigg[
(m^2_{h} - 2 m^2_{H^\pm})s_{\beta-\alpha} s^2_{2\beta} - 
2c_{\beta + \alpha}(m^2_{h} s_{2\beta}-
2 m^2_{12})\bigg], \label{triple-hhh}
\end{eqnarray}
where $g$ is the $SU(2)$ gauge coupling constant. 
We have used the notation $s_x$ and $c_x$ as short-hand for 
$\sin(x)$ and $\cos(x)$, respectively. 
It is clear from the above couplings that $Hhh$ is proportional to 
$c_{\beta-\alpha}$ which is close to unity in our case and hence the BR$(H\to hh)$
would not be suppressed. {{Moreover, 
in the exact fermiophobic limit $\alpha\approx \pm \pi/2$ 
becomes proportional to $m_{12}^2$.}}
The vertex $HAA$ has two terms, one proportional to 
$c_{\beta-\alpha}$ and the other proportional to $s_{\beta+\alpha}$ which
is close to $c_\beta$ in the fermiophobic limit $\alpha\approx \pi/2$. Finally,
the coupling $hH^\pm H^\mp$  can be large so as to  contribute sizably to the 
$h\to \gamma\gamma$ decay rate.


\subsection*{Fermiophobic limit of the 2HDM-I}
In general, in the 2HDM, both Higgs doublets can couple to quarks and leptons
exactly as in the SM. However, in such case one has tree level FCNCs which would lead to large contribution to $B$-physics
observables in conflict with data. In order to avoid this,
the 2HDM needs to satisfy a discrete $Z_2$ symmetry 
 \cite{Glashow:1976nt,Branco:2011iw} which guarantees the absence of 
this phenomenon. 
Several type of 2HDMs exist depending on the $Z_2$ 
charge assignment of the Higgs doublets~ \cite{Branco:2011iw}. 
In our study, we will focus on the 2HDM-I. In this model
only the doublet $\phi_2$ couples to all 
the fermions as in the SM while $\phi_1$ does not couple to any of the 
fermions. 


The Yukawa interactions in terms of the neutral and charged Higgs 
mass eigenstates in a general 2HDM can be written as:
\begin{eqnarray}
-\mathcal{L}^{2HDM}_{Yukawa}&&=\Sigma_{f=u,d,l}\frac{m_f}{v}(\xi^h_f\bar{f}fh+\xi^H_f\bar{f}fH-i\xi^A_f \bar{f}\gamma_5 fA)\nonumber\\
&&+\{
\frac{\sqrt{2}V_{ud}}{v}\bar{u}(m_u\xi^A_u P_L+m_d \xi^A_d P_R)dH^+ 
+\frac{\sqrt{2}m_l\xi_l^A}{v}\bar{\nu}_L l_RH^++{\rm h.c.}\},
\label{l-yuk}
\end{eqnarray}
where $v^2=v_1^2+v_2^2=(2\sqrt{2}G_F)^{-1}$, $V_{ud}$ is the top-left entry of
the Cabibbo-Kobayashi-Maskawa (CKM) matrix and $P_L$ and $P_R$ are the left-
and right-handed projection operators, respectively. In the 2HDM-I, we have
\begin{eqnarray}
&& \xi^h_f=\cos\alpha/\sin\beta \quad  and \quad  
\xi^H_f=\sin\alpha/\sin\beta , \quad for \quad
f=u,d,l, \nonumber\\ 
&& \xi^A_d=-\cot\beta \quad , \quad \xi^A_u=\cot\beta \quad and \quad 
\xi^A_l=-\cot\beta. 
\label{HhA-coup}
\end{eqnarray}

From the above Lagrangian~(\ref{l-yuk}) and (\ref{HhA-coup}), 
it is clear that for $\alpha\approx \pm\frac{\pi}{2}$ the tree 
level coupling of the 
light CP-even Higgs $h$ to  quarks and leptons are very suppressed. 
Hence $h$ is fermiophobic in this limit~ \cite{Akeroyd:1995hg}.
Note that, in the 2HDM-I, the CP-odd Higgs coupling to fermions 
is proportional to 
$\cot\beta$ and hence would not be fermiophobic for any choice of $\tan\beta$.
Since we are interested in the case where $H$ is SM-like
($\cos(\beta-\alpha)\approx 1$) and  $m_h\leq m_H/2 \approx 62.5$ GeV 
such that the decay $H\to hh$ is open, the main decays of the lightest  
Higgs state  $h$ are into the tree level channels $h\to V^*V^*$ and 
$h\to Z^* A$ when $m_A<m_h$, otherwise the one loop $h\to Z^*\gamma$ and 
$h\to \gamma \gamma$ ones dominate.   
The $1\to 4$ decays $h\to V^*V^*(\to f\bar f f'\bar f')$ have two 
sources of suppression, 
the  phase space one and the fact that $hVV\propto \sin(\beta-\alpha)\approx
0$, while  the $1\to 3$ decay $h\to Z^*\gamma(\to f\bar f\gamma)$ 
is both loop and phase space suppressed. 
Therefore, the decays $h\to \gamma \gamma$ and $h\to Z^*A$ ($m_A<m_h$) are 
expected to compete with each other and dominate in the fermiophobic limit.
In fact, it is well known than  $h\to \gamma\gamma$ is dominated by the $W^\pm$ 
loops which interfere destructively with top and charged Higgs loops.
In the limit where $\sin(\beta-\alpha)\to 0$, 
the $W^\pm$ loops vanish and only the top and charged Higgs  ones contribute. 
When $\cos\alpha$ vanishes, the $h$ state, with mass $\leq 62$ GeV, 
becomes fermiophobic and consequently the  
{\rm BR}($h\to \gamma\gamma$) can become 100\% if $h\to Z^*A$ is not open. 
In contrast, the coupling $hZA$ is proportional to $\cos(\beta-\alpha)$,
which is close to unity in our scenario, therefore, when $h\to Z^*A$ is open
for $m_A<m_h$, it dominates over $h\to \gamma\gamma$.

Fermiophobic Higgs bosons have been searched for at LEP 
and Tevatron.  The LEP collaborations used
$e^+e^-\to Z^* \to Zh$ followed by the decay $h\to \gamma\gamma$
and set a  lower limit on $m_h$ of the order $100$ GeV 
 \cite{Abbiendi:2002yc, Abreu:2001ib,Heister:2002ub,Achard:2002jh}.
At the Tevatron, both Higgs-strahlung ($pp\to Vh$, $V=W^\pm,Z$) 
and vector boson fusion ($qq\to q'q'h$)  have been used
to search for fermiophobic Higgs decays of the type 
$h\to \gamma\gamma$  \cite{Abazov:2008ac}, with 
similar results to those obtained at LEP. 
Note that both LEP and Tevatron assumed a full SM coupling for $hVV$ ($V=Z,W$)
which  would not be the case for the CP-even Higgs $h$ in the 2HDM-I  
where $hVV\propto \sin(\beta-\alpha)$ can be very suppressed, as explained.
Therefore one could imagine a scenario with a very light $h$ state
($m_{h}\ll 60$ GeV) which has escaped  LEP and  
Tevatron limits due to suppression in the coupling $hVV$.
In addition, the LEP, OPAL and DELPHI collaborations have searched for 
fermiophobic Higgs decays through $e^+e^-\to Z^*\to Ah$ with 
$h\to \gamma\gamma$ and $A$
decaying mainly into fermions and set a limit on 
$\sigma(e^+e^-\to Ah)\times {\rm BR}(h\to
\gamma\gamma)\times {\rm BR}(A\to f\bar{f})$ for $m_h\in [20,180]$ 
GeV  \cite{Abbiendi:2002yc,Abreu:2001ib}. Note that 
this limit will depend on the coupling $ZhA\propto \cos(\beta-\alpha)$
and hence becomes weaker for $\cos(\beta-\alpha)\ll 1$.
However, a very light $h$ with  $m_h\leq 60$ GeV 
is sill allowed if the CP-odd is rather heavy.  
We refer to  \cite{Arhrib:2017wmo} for more detail on these aspects.
Finally, following phenomenological studies in  \cite{Akeroyd:2005pr},
CDF at Tevatron also studied $qq'\to H^\pm h$, which can lead to 4-photon final
states for $H^\pm \to W^\pm h$ and $h\to \gamma\gamma$  \cite{Aaltonen:2016fnw},
however, CDF limits are presented only for the exactly fermiophobic
scenario and are not readily extendable to our more general setup.

\section{Numerical results}
As said previously, we are interested in the 2HDM-I 
for which we perform a systematic numerical scan for its parameter space. 
We have fixed $m_H$ to 125 GeV and assumed that $2 m_h<m_H$  such that 
the decay $H\to hh$ is open. The other 2HDM independent parameters 
are varied as indicated in Tab.~1.
We use the 2HDMC (v1.7.0)~ \cite{Eriksson:2009ws} public program 
to calculate the 2HDM spectrum as well as various decay rates and 
BRs of Higgs particles. The 
2HDMC program also allow us to check several theoretical 
constraints such as perturbative unitarity, 
boundedness from below of the scalar potential 
as well as EW Precision Observables (EWPOs) which are all 
turned on during the scan. 
In fact, it is well known that EWPOs constrain the splitting 
between Higgs masses. In our scenario, since we ask that $m_H=125$ 
GeV and assume $2m_h<m_H$,   if we want to keep the CP-odd also light, 
it turns out that the charged Higgs boson 
would be also rather light, $m_{H^\pm}\leq 170-200$ GeV  \cite{Enberg:2016ygw},
as it can be seen from Tab.~1.
Moreover, the code is also linked to  
HiggsBounds \cite{Bechtle:2013wla} and
HiggsSignals \cite{Bechtle:2013xfa}
that allow us to check various LEP, 
Tevatron and recent LHC searches. 

Once in the 2HDM-I the decay channels $H\to hh$ and/or $H\to AA$ are open, the 
subsequent decays of $h$ and/or $A$ into fermions, photons or gluons will  
lead either to invisible $H$ decays that can be constrained by
 present ATLAS and CMS data on the Higgs couplings. 
In our study, we will use the fact that the total BR of
the SM-like Higgs boson into undetected BSM decay modes is constrained, 
as mentioned, by BR$_{\rm BSM}< 0.34$  \cite{brbsm} where 
BR$_{\rm BSM}$ will designate in our case the sum of BR$(H\to hh)$ and 
BR$(H\to AA)$.

In what follow, we will show our numerical results via  three 
different scans (see Tab.~1). 
These results mainly concern the BR$_{\rm BSM}$, BR$(H\to hh)$ and the ensuing 
total cross section for four photons final states which is given by 
\begin{eqnarray}
\sigma_{4\gamma}&=& 
\sigma(gg\to H)\times {\rm BR}(H\to hh)\times {\rm BR}^2(h\to \gamma\gamma).
\label{eq:4gamma}
\end{eqnarray}
Note that, in writing the above cross section,  we have used the narrow width 
approximation for the SM-like Higgs state $H$ which is justified since 
the total width of $H$ is of the order of few MeV (see Tab.~3). Furthermore, 
 we are interested into multi-photon signatures coming from 
$H\to hh\to 4\gamma$  and not from $H\to AA\to 4\gamma$ decay. 
In the former case,  in fact, because $h$ can become totally fermiophobic,  
its  BR$(h\to \gamma\gamma)$ can in turn become maximal when $h\to 
Z^*A$ is not open.  In the latter case, of the CP-odd Higgs state $A$,  
couplings to fermions are proportional to
$1/\tan\beta$,  which thus does not vanish, 
therefore the one loop decay $A\to \gamma\gamma$ 
will be suppressed compared to the tree level ones 
$A\to ff$ and $A\to Z^*h$. 

\begin{figure}[h]
\centering
\includegraphics[width=0.45\textwidth,height=0.4\textwidth]{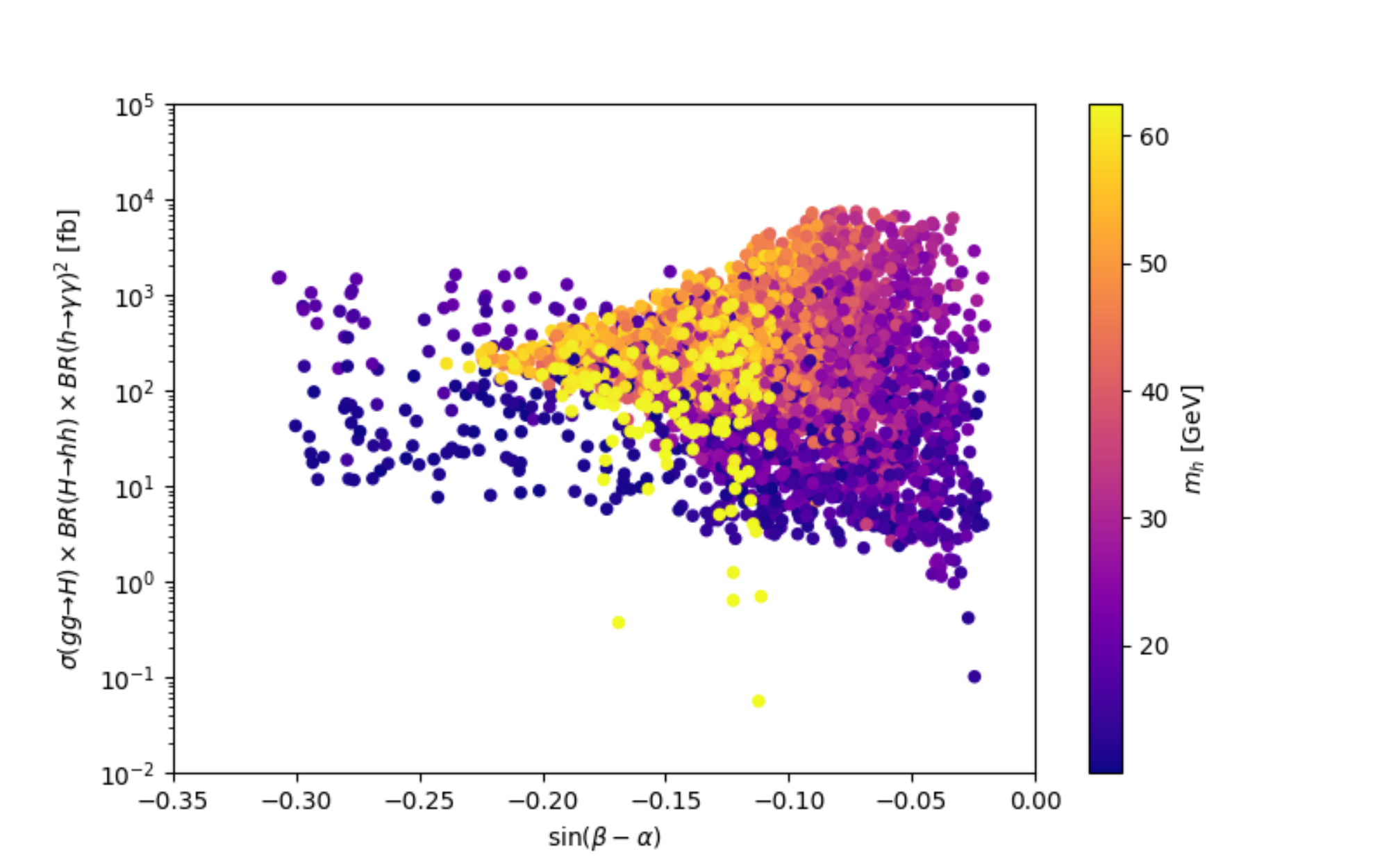}
\includegraphics[width=0.45\textwidth,height=0.4\textwidth]{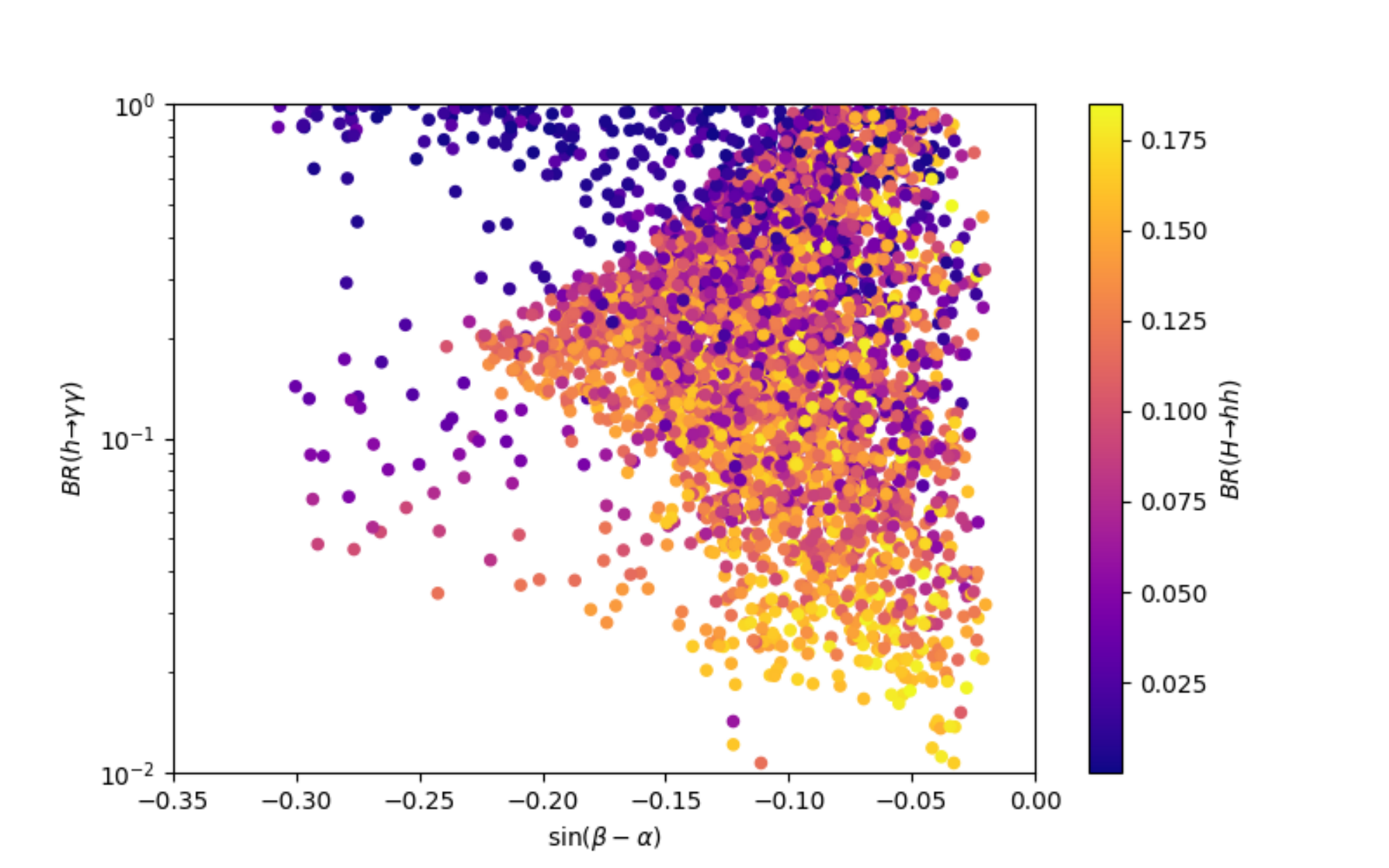}
\caption{(Left) The $\sigma_{4\gamma}$ rate as a function of $\sin(\beta-\alpha)$
  with $m_h$  indicated on the right vertical axis. (Right) The ${\rm
    BR}(h\to \gamma\gamma)$ as a function of $\sin(\beta-\alpha)$ with
  ${\rm BR}(H\to hh)$ indicated on the right vertical axis. Both plots are for scan-1.}
\label{scan1-fig2}
\end{figure}

 We first show our results for $\sigma_{4\gamma}$ without imposing 
 constraint from ATLAS searches in events with at least three photons in the
 final state \cite{beta}.
The results of scan-1 are shown in Fig.~\ref{scan1-fig2}. In this scan
we only allow $H\to hh$ to be open and deviate from the exact 
fermiophobic limit by taking
$\alpha=\pm\pi/2 \mp \delta$ where $\delta\in [0,0.05]$. 
It is clear that for some values of $\delta \approx 0$, 
one can have an exact fermiophobic Higgs with maximal BR 
for $h\to \gamma\gamma$. In this scenario, 
The BR$(H\to hh)$ can reach 17\% in some cases.
Thus, the four photon cross sections can 
become of the order of few pb when 
{\rm BR}($h\to \gamma\gamma$) is close to maximal and BR$(H\to hh)$ large. 
Here, the maximum cross section is reached for
$\sin(\beta-\alpha)\approx -0.06$.

\begin{figure}[h]
\centering
\includegraphics[width=0.45\textwidth,height=0.4\textwidth]{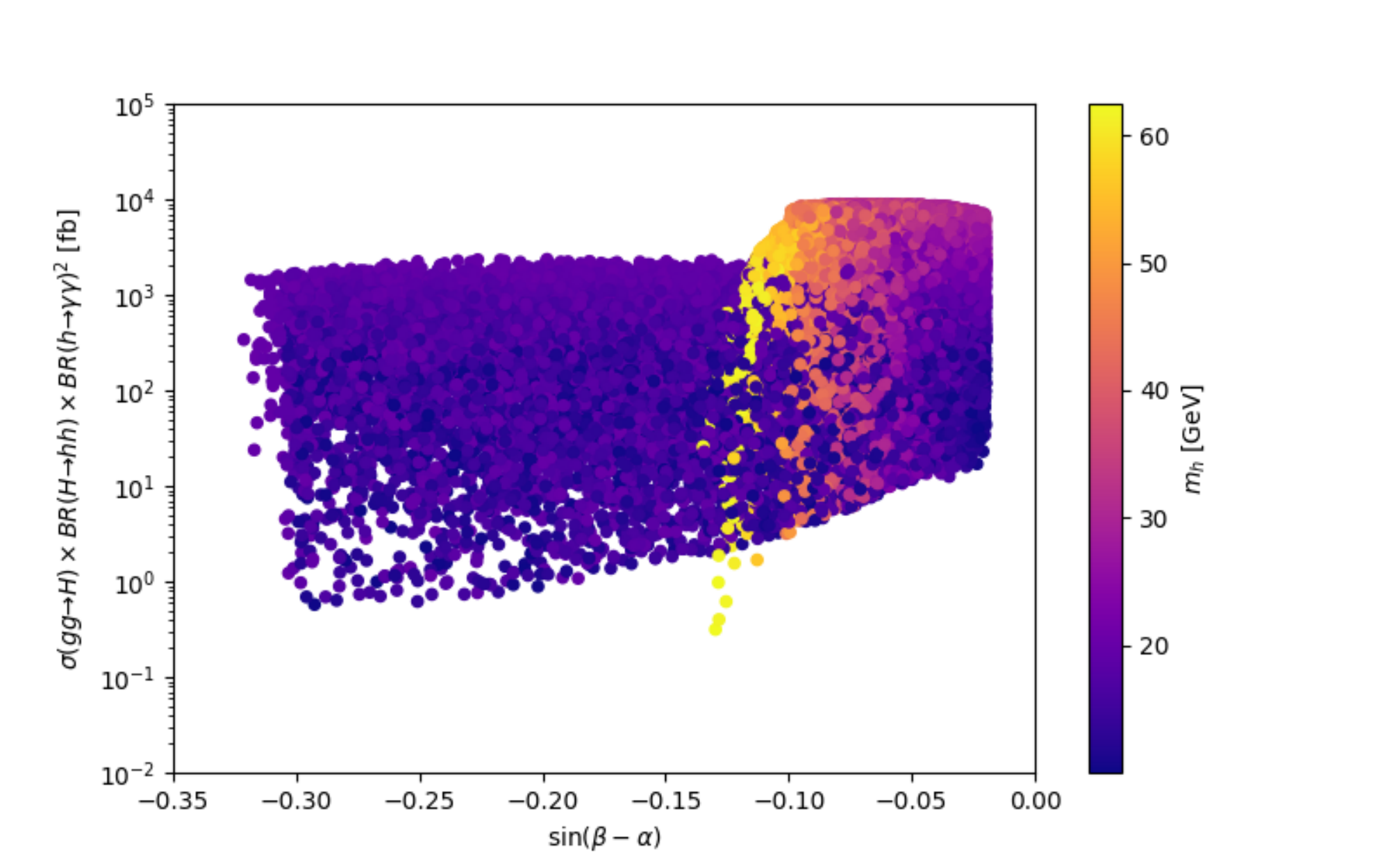}
\includegraphics[width=0.45\textwidth,height=0.4\textwidth]{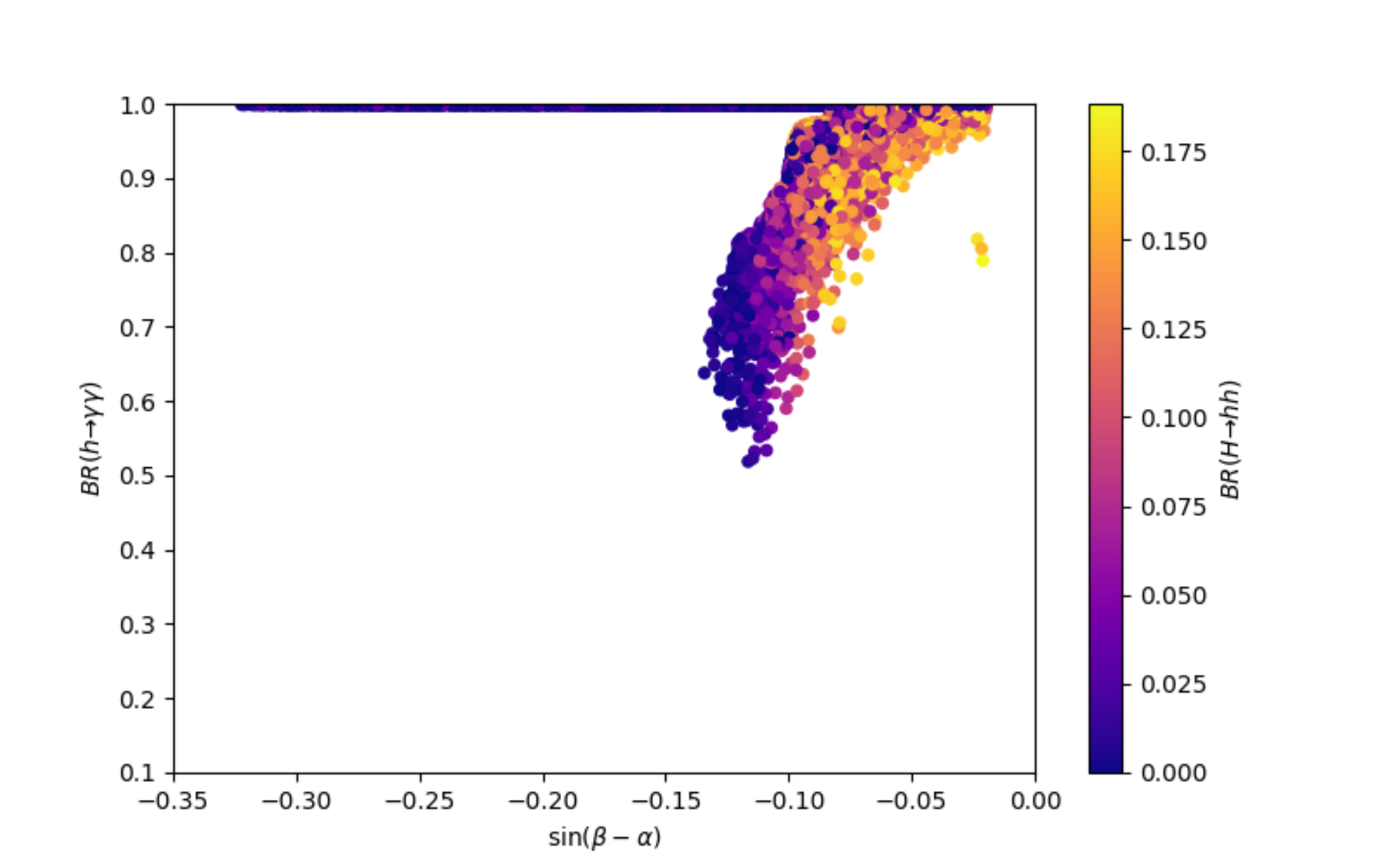}
\caption{(Left)  The $\sigma_{4\gamma}$ rate as a function of $\sin(\beta-\alpha)$
  with $m_h$ indicated on the right vertical axis. (Right) The ${\rm
    BR}(h\to \gamma\gamma)$ as a function of $\sin(\beta-\alpha)$ with 
${\rm BR}(H\to hh)$  indicated
  on the right vertical axis. Both plots are for scan-2.}
\label{scan2-fig3}
\end{figure}

{{The output of scan-2, which is for the exact fermiophobic limit, 
$\alpha=\pi/2$, is shown in Fig.~\ref{scan2-fig3}.  Here, 
we illustrate  $\sigma_{4\gamma}$ as a function of $\sin(\beta-\alpha)$ 
in the left panel with $m_h$ coded with different colours on the vertical axis.
The BR$(h\to \gamma\gamma)$ as a function of $\sin(\beta-\alpha)$ is depicted  
on the right panel of  Fig.~\ref{scan2-fig3} with the BR$(H\to hh)$ on the
vertical axis. The maximal value reached by BR$(H\to hh)$ in this scenario is again around 17\%.
Note that, in this case of exact fermiophobic limit, only $W^\pm$ and $H^\pm$ loops
contribute to the $h\to \gamma\gamma$ decay. In fact, in most cases,
$W^\pm$ loop contributions to $h\to \gamma\gamma$ dominate over 
 the $H^\pm$ ones except for small $\sin(\beta-\alpha)$, where 
$W^\pm$ and $H^\pm$ terms could become comparable and interfere destructively. 
In such a case, it may be possible that BR($h\to \gamma\gamma$) will be
suppressed and BR($h\to W^*W^*$) slightly enhanced.
This could explain the drop of the BR($h\to \gamma\gamma$) up to 
$5.5\times 10^{-1}$. Note that, for large $\sin(\beta-\alpha)\approx -0.14$, the
off-shell decay $h\to V^*V^*$ can reach $3.5\times 10^{-1}$, $1\times 10^{-1}$ for $V = W$ and $V = Z$, respectively.
It is interesting to see that, in this scenario, $\sigma_{4\gamma}$ can be
larger than 1 pb for a light $m_h\in [10,50]$ GeV with significant BR$(h\to
\gamma\gamma)$.  
}}

\begin{figure}[h]
	\centering
	\includegraphics[width=0.45\textwidth,height=0.4\textwidth]{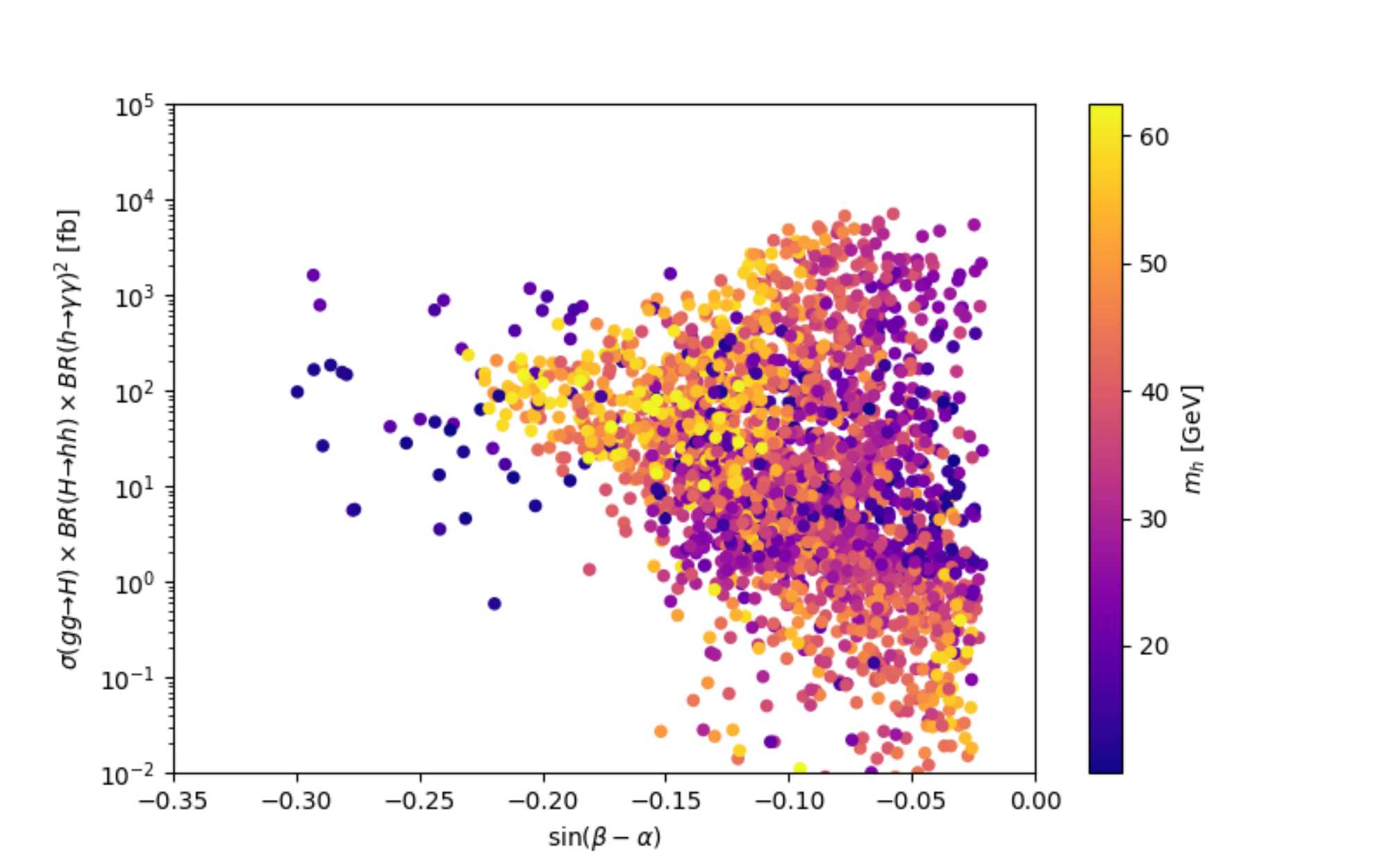}
	\includegraphics[width=0.45\textwidth,height=0.4\textwidth]{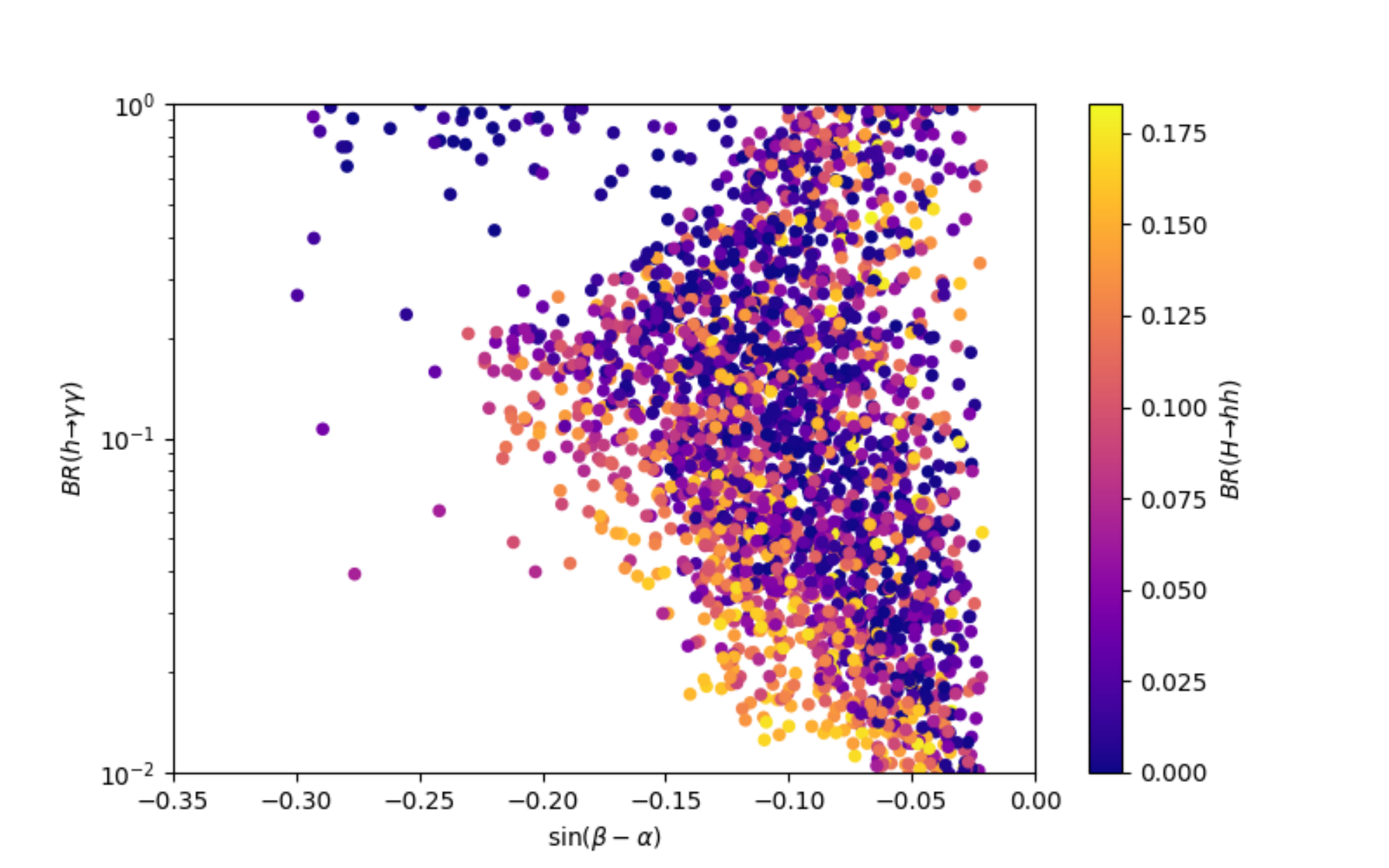}
	\caption{(Left) The   $\sigma_{4\gamma}$ rate as a function of
          $\sin(\beta-\alpha)$ with $m_h$  indicated on the right vertical
          axis. (Right) The ${\rm BR}(h\to \gamma\gamma)$ as a function of
          $\sin(\beta-\alpha)$ with ${\rm BR}(H\to hh)$  indicated on 
          the right vertical axis. Both plots are for scan-3.}
	\label{scan3-fig4}
\end{figure}

{{
In scan-3, we allow $\sin(\beta-\alpha)\in [-0.35,0]$  and  
the CP-odd Higgs state to be as light as 10 GeV. In this case, 
we specifically take into account constraints from the 
LEP measurement of the $Z$ width. The results are illustrated in 
Fig.~\ref{scan3-fig4}, where we show both $\sigma_{4\gamma}$ and 
${\rm BR}(h\to \gamma\gamma)$ as a function of $\sin(\beta-\alpha)$. 
Note that, for any choice of $\tan\beta$, one can tune  
$\sin(\beta-\alpha)$ such that $\alpha$  becomes $\pm\pi/2$ and then $h$ 
is fermiophobic (in which case the previous discussion of 
scan-2 would apply again). Away from this fermiophobic limit, 
BR$(h\to b\bar{b})$ becomes sizeable and suppresses 
${\rm BR}(h\to \gamma\gamma)$.
}}

{{
After quantifying the maximal size of the four photons cross section in the
previous plots, we proceed to apply 
ATLAS limits coming from searches in events with at least 
 three photons in the final state \cite{beta}. The results are shown in 
Fig.~\ref{fig4}. 
The solid black line is the expected upper limit at 95$\%$ CL from ATLAS with 
8 TeV centre-of-mass energy and 20.3 fb$^{-1}$ luminosity. 
The green and yellow bands 
 correspond, respectively, to a $\pm 1$ $\sigma$ and 
$\pm 2$ $\sigma$ uncertainty from the resonance search assumption. 
As it can be seen from this plot, for $m_h$ in the
 $[10,62]$ GeV range, the ATLAS upper limit on 
$\sigma_H\times BR(H\to hh)\times {\rm BR}^2(h\to
\gamma\gamma)$ is $1\times 10^{-3}\sigma_{SM}$.
We also illustrate on this figure our projection for 14 TeV using
 300 fb$^{-1}$  luminosity (based on a MC simulation that we will describe below).
The  dots represent our surviving points from scan-1, scan-2 and scan-3 
after passing  all theoretical and experimental constraints. 
Most of the points with significant four photons cross section and large ${\rm BR}(H\to hh)$ and/or ${\rm BR}(h\to \gamma\gamma)$ 
shown in the previous plots turn out to be ruled out by the aforementioned ATLAS upper 
limit \cite{beta}.  It is clear from Fig.~\ref{fig4} (top-left and -right plots) that 
scenarios from scan-1 and scan-2 would be completely ruled out (or, conversely, be discovered) by our
projection for the 14 TeV LHC run with 300 fb$^{-1}$ luminosity while 
scenarios from scan-3 (bottom plot) would survive undetected. 
The maximal four photons cross section we obtain is
of the order of 37 fb. It is interesting to note that, for scan-1 and scan-3, the remaining points
 still enjoy sizable BR$(H\to hh)$ while for the exact fermiophobic limit of 
 scan-2 one can see from Fig.~\ref{fig4} (top-right) that the BR$(H\to hh)$
 is less than $5\times 10^{-3}$. This limit is much stronger than the one 
from invisible SM-like Higgs decays discussed previously.
This can be seen in Fig.~\ref{fig5}, where we illustrate
 the correlation between ${\rm BR}(H\to \gamma\gamma)$ and 
${\rm BR}(H\to hh)$ for the three scans. Herein, one can verify  that, for scan-1 and
 scan-3, ${\rm BR}(H\to \gamma\gamma)$ and  ${\rm BR}(H\to hh)$ are
 anti-correlated. }}

Based on the results of these three scans we have selected a few 
Benchmark Points (BPs),  which are given in Tab.~2. These BPs can be seen
in Fig.~\ref{fig4} as black stars.
Note that, in BP1, both $H\to hh$ and
$H\to AA$ decays are open while for the other BPs only $H\to hh$ is.
For these BPs, we give in Tab.~2 various observables 
such as: the total widths of $h$ and $H$, $\Gamma_{h}$ and $\Gamma_H$,  
respectively,  ${\rm BR}(H\to hh)$, 
${\rm BR}(h\to \gamma\gamma)$, ${\rm BR}(h\to Z^*A)$, 
${\rm BR}(A\to \gamma\gamma)$  and the 
four photons cross section $\sigma_{4\gamma}$ in fb. 
In fact, for these BPs, we take into account all theoretical 
constraints as well as the 
LEP and LHC constraints implemented in the HiggsBounds code plus the  
limits from
ATLAS on multi-photons final states  \cite{beta}, as explained in the 
introduction, {{see Eqs. (1) and (2)}}.
It is also interesting to see from Tab.~3 that the 
${\rm BR}(A\to \gamma\gamma)$ is
always suppressed and cannot be used to generate multi-photon finale
states. Finally, it  can also  be  seen from this table that, 
even for small ${\rm BR}(H\to
hh)\approx 10^{-3}$ but with maximal ${\rm BR}(h\to \gamma\gamma)$,  
one can still get a large $\sigma_{4\gamma}\approx 38$ fb.

Before ending this section, we would like to comment on charged Higgs and
CP-odd Higgs boson  searches. As mentioned previously, 
the charged Higgs and CP-odd Higgs states are rather light in our scenarios. 
LHC limits on light charged Higgs states produced from top decay and 
decaying to $H\pm \to \tau\nu, cb$ in the 2HDM-I can be evaded
by advocating the dominance of the $H^\pm \to W^\pm A$ or $H^\pm \to W^\pm h$
BRs (see  \cite{Arhrib:2017wmo} for more details). 
On the one hand, the LHC searched for a CP-odd Higgs state decaying 
via  $A\to ZH$  \cite{Aad:2015wra,Khachatryan:2015tha,Khachatryan:2015lba} 
and $A\to\tau^+\tau^-$. In our scenario, the 
BR$(A\to ZH$) will suffer two
suppressions: one coming from the coupling 
$AZH$, which is proportional to $\sin(\beta-\alpha)\approx 0$, 
and the other one coming from the fact that $A\to Zh$ would dominate over 
$A\to ZH$ since $h$ is lighter than 125 GeV and the coupling $ZAh$ is 
proportional to $\cos(\beta-\alpha)\approx 1$. On the other hand,  ATLAS
and CMS searches for a CP-odd Higgs state decaying to a pair of $\tau$ leptons
 \cite{Aad:2014vgg,Khachatryan:2014wca}, when applied to the 2HDM-I, only
exclude small $\tan\beta\leq 1.5$ for $m_A\in [110,350]$ 
GeV  \cite{Arhrib:2015gra}. This can be understood easily from the fact that 
$A$ couplings to a pair of fermions in the 2HDM-I are proportional to
$1/\tan\beta$, hence both the production $gg\to A$ and the 
decay $A\to \tau^+\tau^-$ are suppressed for large $\tan\beta$ values. Moreover,
in our scenario, BR$(A\to \tau^+\tau^-)$ would receive an other 
suppression from  the opening of the $A\to Z^* h$ channel. 
Note also that LEP limits on a light $h$ and a light $A$ are implemented 
in the HiggsBounds code through limits on  the processes $e^+e^- \to Zh$ and 
$e^+e^- \to hA$.


\begin{table}[hpbt]
	\begin{ruledtabular}
		\begin{tabular}{c c c c }
			\hline parameters  & scan-1 & scan-2 & scan-3\\
			\hline $m_H$ (SM-like)  & 125&  125 & 125\\
			$m_h$ & $[10,62.5]$ & $[10,62.5]$ & $[10,62.5]$\\
			$m_A$  & $[62.5,200]$& $[62.5,200]$ & $[10,200]$\\
			$m_{H^\pm}$     & $[100,170]$&$[100,170]$& $[100,170]$\\
			$\tan\beta$ & $[2 , 50]$ & $[2 , 50]$ & $[2,50]$\\
		    $\alpha$  & $\alpha$=$\pm\frac{\pi}{2} \mp\delta $& $\alpha$=$\frac{\pi}{2}$ & $s_{\beta-\alpha}=[-0.35,0.0]$\\
			$m_{12}^2$   & $[0,100]$  &$[0,100]$& $[0,100]$ \\
			$\lambda_6=\lambda_7$  & 0 & 0 & 0\\
			\hline 
		\end{tabular}
		\caption{2HDM parameters scans: all masses are in GeV.}
	\end{ruledtabular}
\end{table}


\begin{table}[h]
	\begin{ruledtabular}
		\begin{tabular}{c | c c c c c c c c}
			\hline 
			& $m_h$ & $m_A$& $m_{H^\pm}$ & $\sin(\beta-\alpha)$ & $\tan\beta$ & $m^2_{12}$ & $\Gamma_{h}$ & $\Gamma_{H}$\\
			\hline 
			BP1 & 10.744652 & 78.567614 & 104.864345 & -0.208633 & 4.584063 & 15.802484 &  $2.21\times 10^{-9}$ & $4.679\times 10^{-3}$\\
			\hline 
			BP2 & 57.440184 & 141.121784 & 116.073489 & -0.114739 & 8.650594 & 9.735405&  $5.303\times 10^{-9}$ & $4.376\times 10^{-3}$\\ 
			\hline 
			BP3 & 40.663472 & 121.812799 & 161.021149 & -0.091551 & 11.262490 & 22.299875&  $1.369\times 10^{-8}$ & $4.507\times 10^{-3}$\\
			\hline 
		\end{tabular}                                                                                      
\caption{Input parameters and widths corresponding to the selected BPs. All
  masses and widths are in GeV and for all points $m_H=125$ GeV.}
	\end{ruledtabular}
\end{table}
\begin{table}[h]
	\begin{ruledtabular}
		\begin{tabular}{c | c c c c c c c  c}
			\hline 
			& ${\rm BR}(h\to \gamma\gamma)$ & ${\rm BR}(H\to hh)$ & ${\rm BR}(A\to \gamma\gamma)$& ${\rm BR}(h\to ZA)$ & ${\rm BR}(A\to Zh)$& $\Gamma_{Z\to A h} [{\rm MeV}]$ &$\beta$  & $\sigma_{4\gamma}$ [fb]\\
			\hline 
			BP1 & 0.1215 & 0.0469 & $7.613\times 10^{-5}$ & 0.0000 & 0.5696 & 0.188995& 0.000694 & 36.92100\\
			\hline 
			BP2 & 0.7435 & 0.001257 & $2.07\times 10^{-5}$ & 0.0000 & 0.9488 & 0.0000& 0.000696 & 35.85200\\ 
			\hline 
			BP3 & 0.1427 & 0.03348 & $1.23\times 10^{-5}$ & 0.0000 & 0.9709 & 0.0000 & 0.000682 & 34.93500\\
			\hline 
		\end{tabular}                                                                                      
\caption{Input parameters, BRs of CP-even and CP-odd Higgs bosons, $Z$ boson
  width and four photon cross section corresponding to the selected
  BPs. All widths are in MeV and for all points $m_H=125$ GeV.} 
\end{ruledtabular}
\end{table}

\begin{table}[h]
	\begin{ruledtabular}
		\begin{tabular}{c | c  c  c  c  c}
			\hline 
			&  Allowed By &  Allowed by theoretical&  Allowed by  &  Allowed by &  Allowed  \\
			& HiggsBounds  &  constraints & HiggsSignals & ATLAS band  & by all constraints\\
			\hline 
			scan-1 & 49.2\% & 20.6\% & 4.5\% & 16.5\% & 0.07\% \\ 
			\hline 
			scan-2 & 81.6\% & 12.5\% & 61.6\% & 5\% & 0.3\% \\ 
			\hline 
			scan-3 & 30,6\% & 2.9\% & 27\% & 63\% & 0.15\% \\
		\end{tabular} 
		\caption{Parameters  as in Tab. III with $10^6$ points as inputs for all scans}. \end{ruledtabular}
\end{table}

\section{Signal and Background}

\begin{figure}[h]
\centering
\includegraphics[width=0.45\textwidth,height=0.4\textwidth]{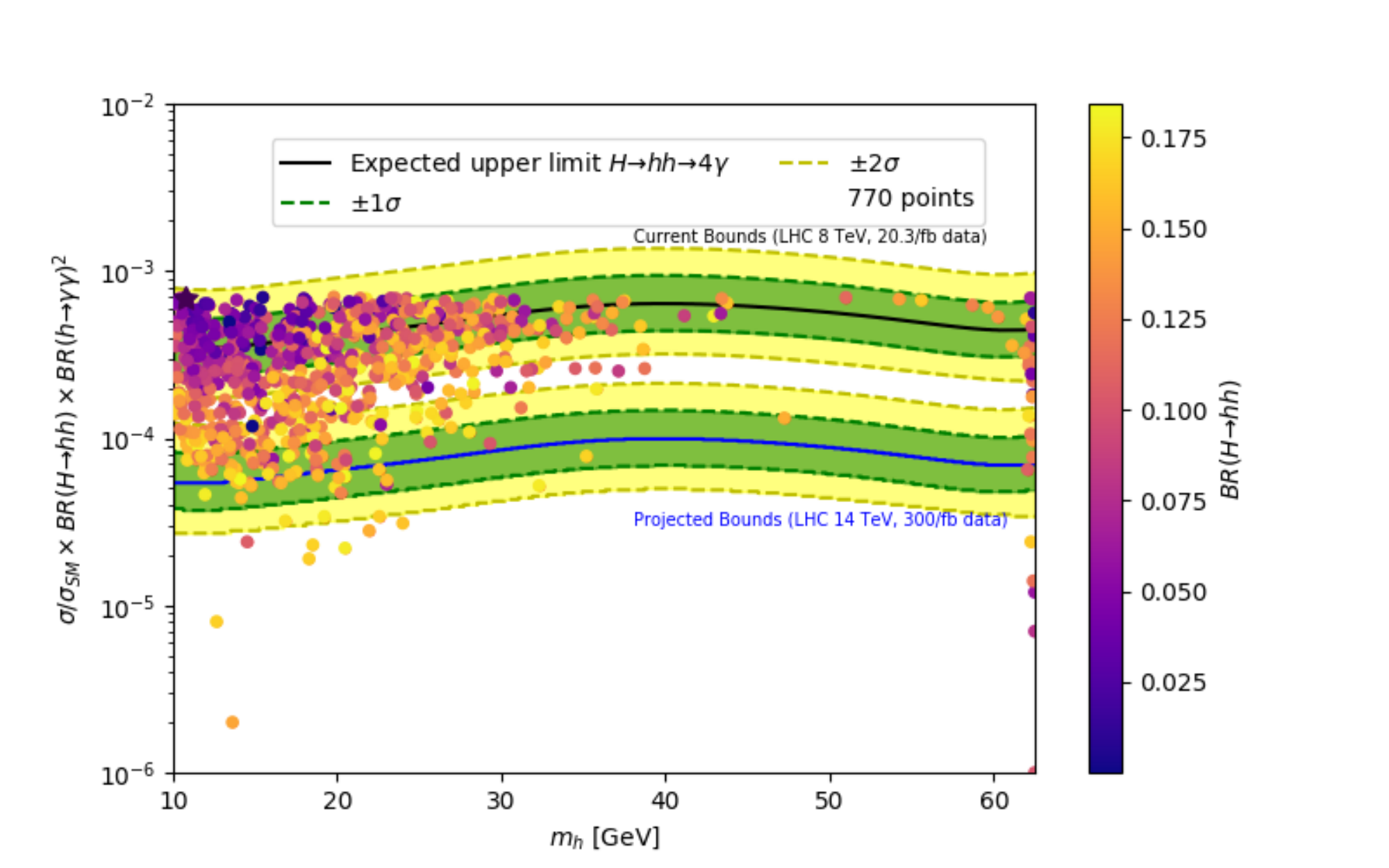}
\includegraphics[width=0.45\textwidth,height=0.4\textwidth]{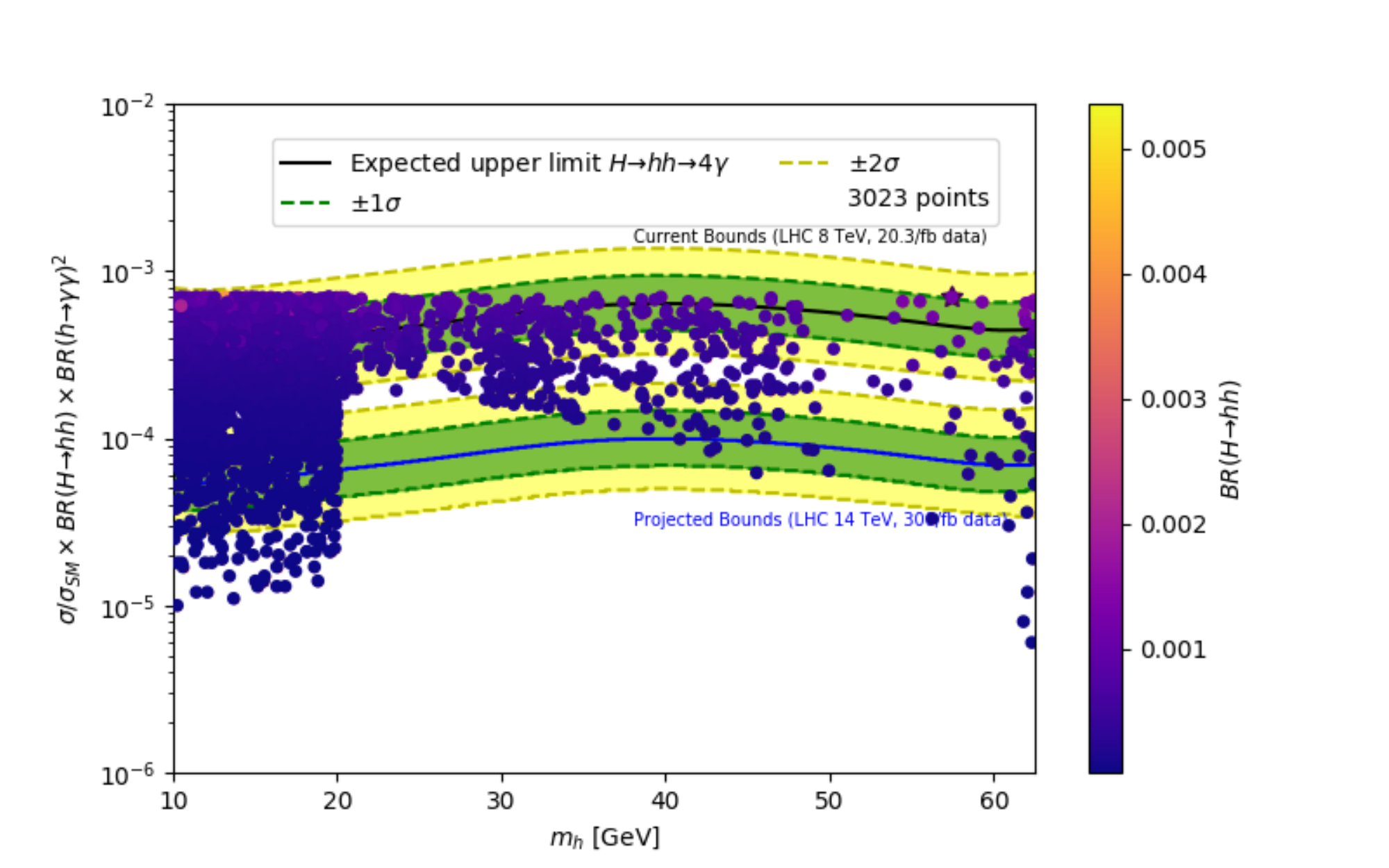}
\includegraphics[width=0.45\textwidth,height=0.4\textwidth]{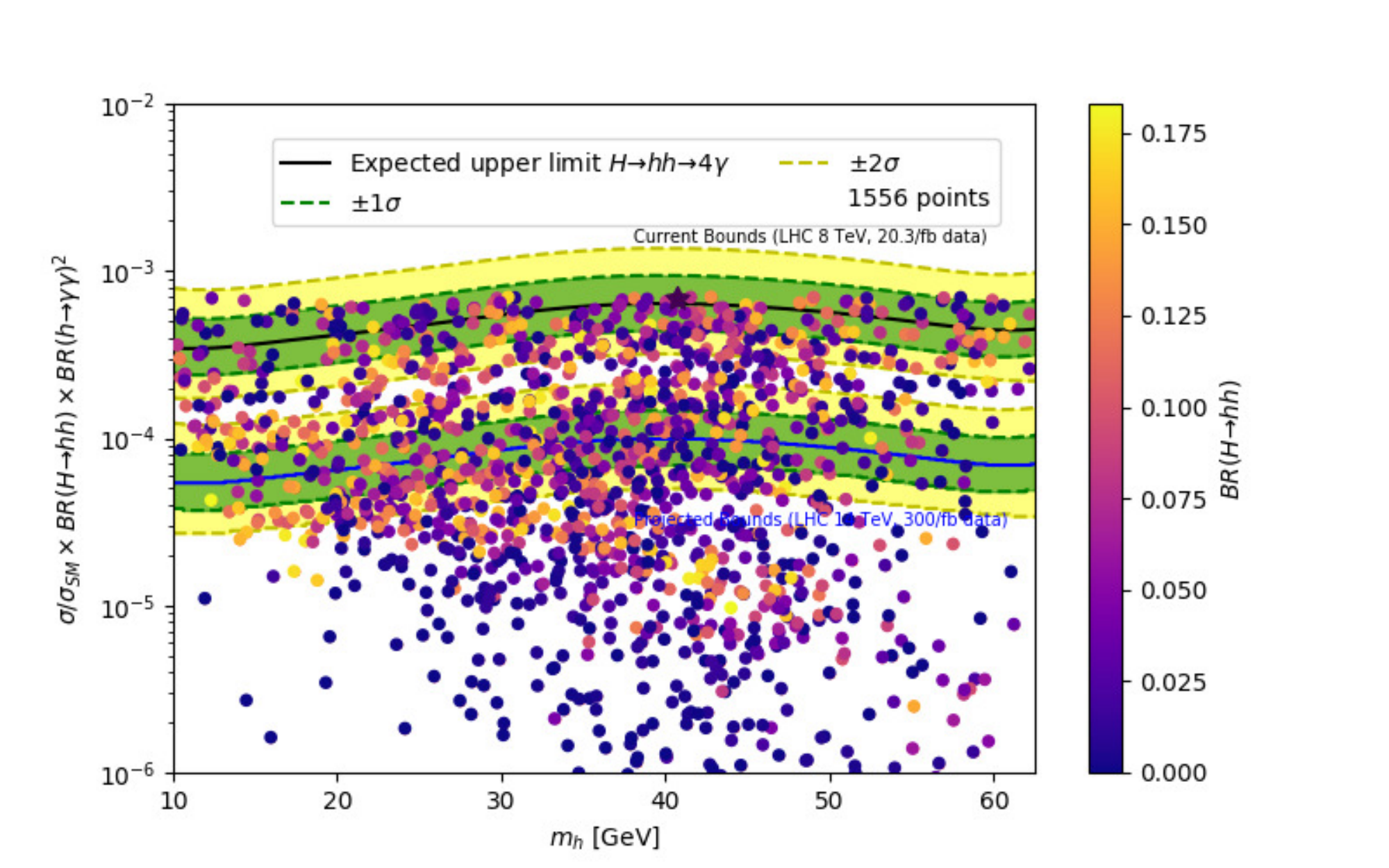}
\caption{Upper limit at 95$\%$ CL on $\sigma_{4\gamma}$ in fb as a
  function of $m_h$ and the $\pm 1$ and $\pm 2$ uncertainty bands resulting from
  ATLAS searches at 8 TeV (upper band) and our projection for 14 TeV (lower band)  
  for (top-left) scan-1,  (top-right) scan-2 and (bottom) scan-3. 
The dots are points that are allowed by all constraints and the black stars 
 represent the BPs given in Tab. II. }
\label{fig4}
\end{figure}

\begin{figure}[h]
\centering
\includegraphics[width=0.45\textwidth,height=0.4\textwidth]{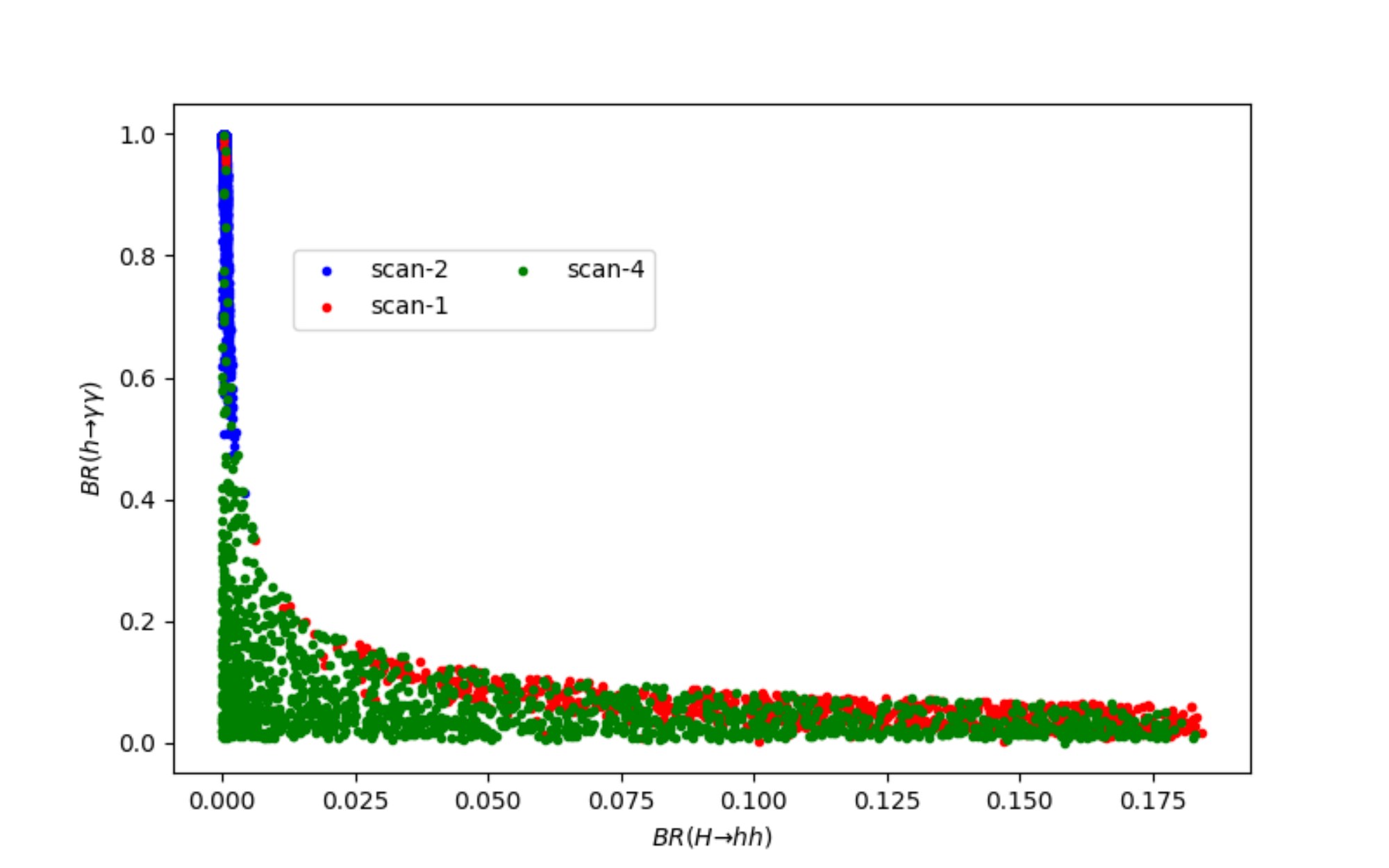}
\caption{{The correlation between BR$(h\to \gamma\gamma)$ and
    BR$(H\to hh)$ in the three scans.}}
\label{fig5}
\end{figure}

{As previously discussed, 
in Figs.~(\ref{fig4}--\ref{fig5}, we have taken into account the constraints from the ATLAS collaboration reported in \cite{atlas} from 8 TeV data. However, in order to project the sensitivity of the future LHC run at $\sqrt{s}=14$ TeV, we have to rescale these results. To determine the `boost factors', for both signal and background processes, needed to achieve this,  we resort to the MC tools.
Specifically, we generate parton level events of both signal and background processes by using MadGraph 5  \cite{Alwall:2014hca} and  then pass them to PYTHIA 6  \cite{Sjostrand:2006za} to simulate showering, hadonisation and  decays. We finally use PGS  \cite{pgs} to perform the fast detector simulations. 
}

In order to pick out the relevant events, for our Run 1 analysis, we adopt the
same selection cuts of the ATLAS collaboration given in   \cite{atlas}, which
read as follows. 
\begin{itemize}
	\item $We assume n_\gamma \geq 3$, i.e., we consider  inclusive three photon
          events.
	\item The two leading  photons should have a $P_t(\gamma) > 22$ GeV
          and  the third one should have a $P_t(\gamma)> 17$ GeV.
	\item The photons should be resolved in the range $|\eta| < 2.37$ and
          do not fall in the endcap region $1.37 < |\eta| < 1.52$.
	\item The cone separation parameter $\Delta R(\gamma \gamma)$ between
          a pair of photons should be larger than $0.4$.
\end{itemize}

One interesting observation is that the kinematics of photons from the
processes $gg \to H \to h h \to 4 \gamma$ and that of $gg \to H \to A A \to 4
\gamma$ are similar when $m_h = m_A$, which could be attributed to the fact
that, although $h$ and $A$ have different parity,  the differential cross
sections of these two processes are both proportional to $(k_1 \cdot k_2)^2 (
k_3 \cdot k_4)^2$ (plus permutations) after the sum over photon polarisations
(the $k_i$'s, for $i=1,2,3,4$, are the photon four-momenta). We provide an
Appendix to demonstrate the details.
\begin{figure}[htbp]
  \centering
  \subfigure{
  \label{Fig1.sub.1}\thesubfigure
  \includegraphics[width=0.44\textwidth]{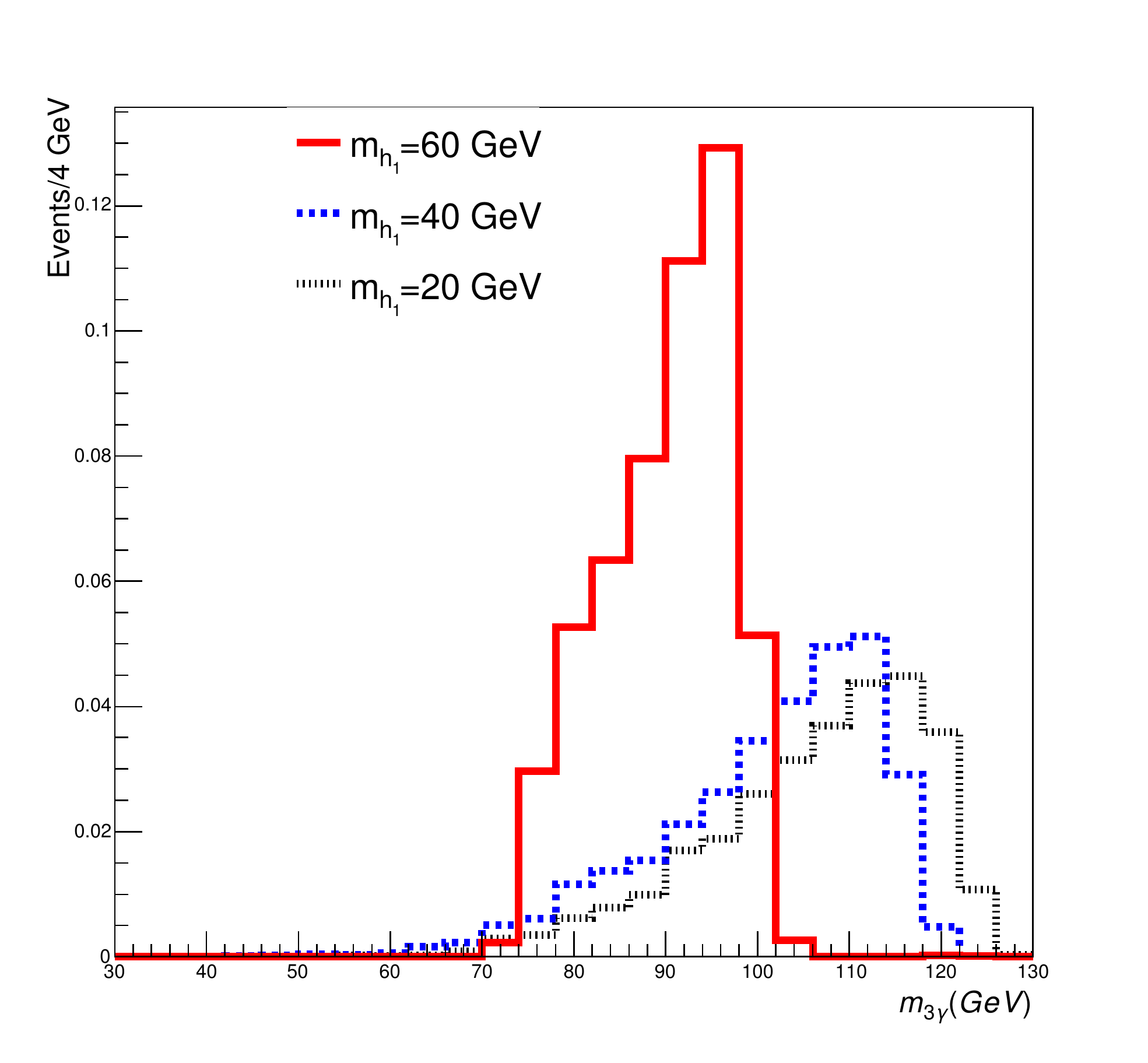}}
  \subfigure{
  \label{Fig1.sub.2}\thesubfigure
  \includegraphics[width=0.44\textwidth]{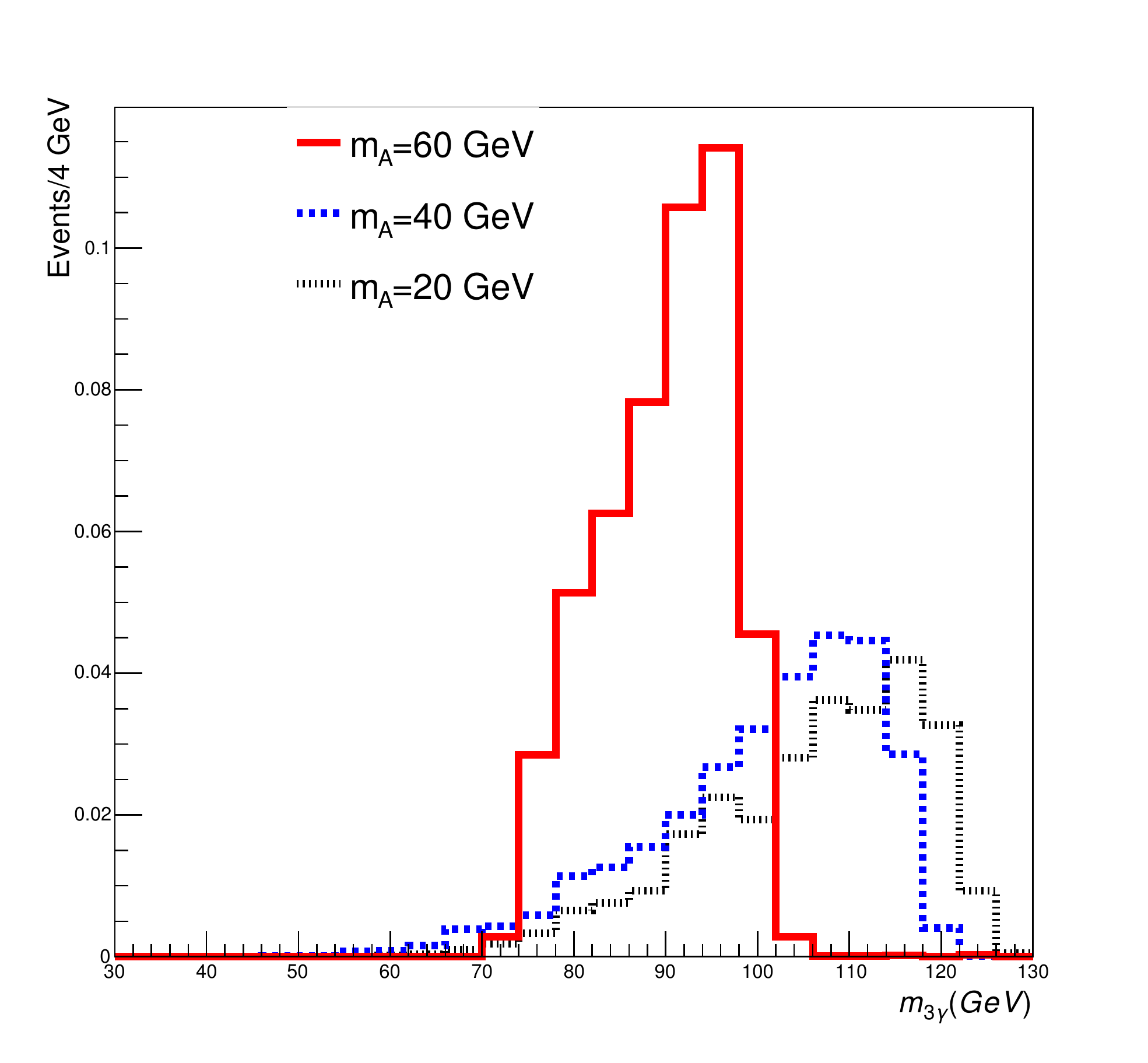}}
  \subfigure{
  \label{Fig1.sub.3}\thesubfigure
  \includegraphics[width=0.44\textwidth]{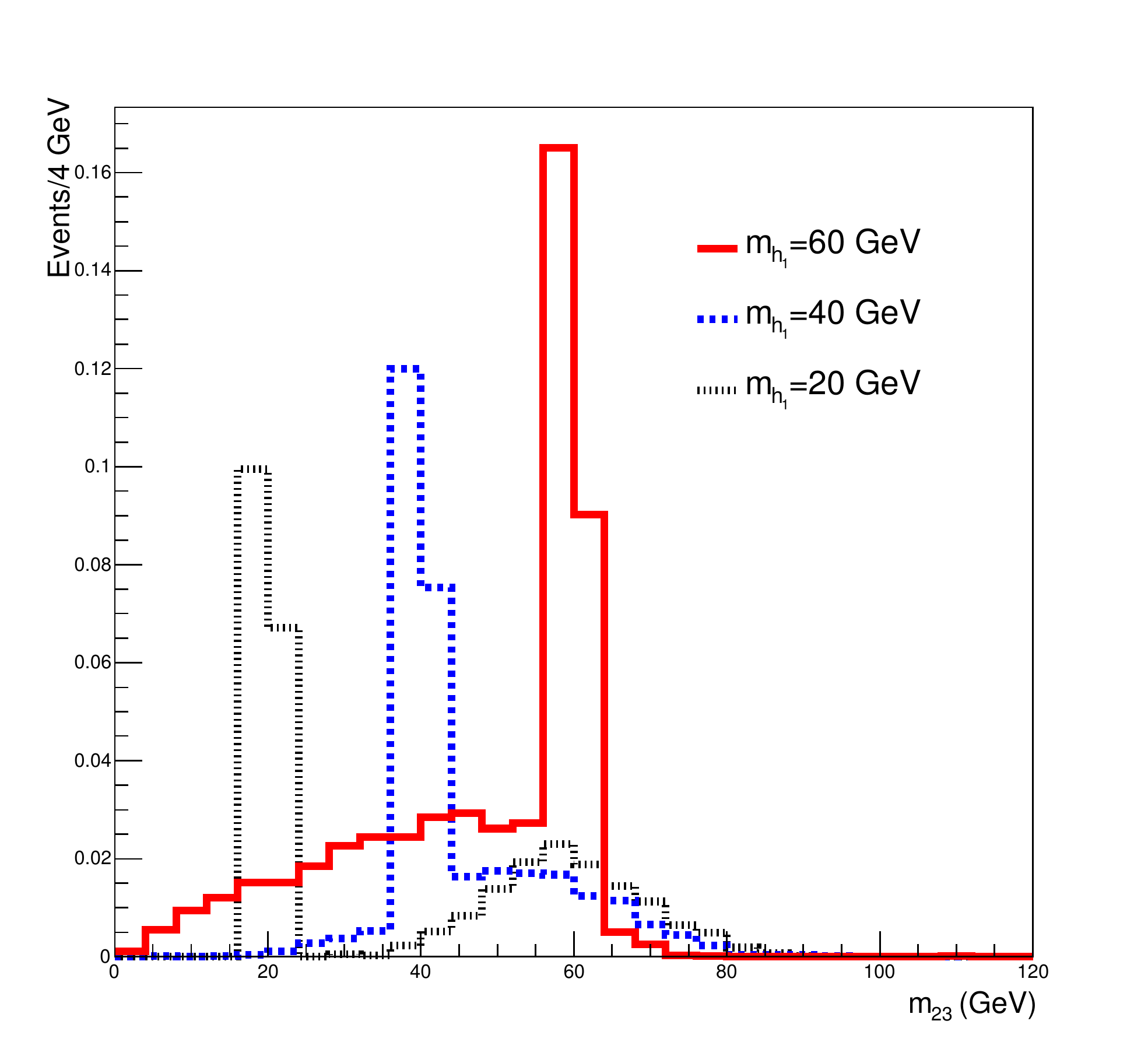}}
  \subfigure{
  \label{Fig1.sub.4}\thesubfigure
  \includegraphics[width=0.44\textwidth]{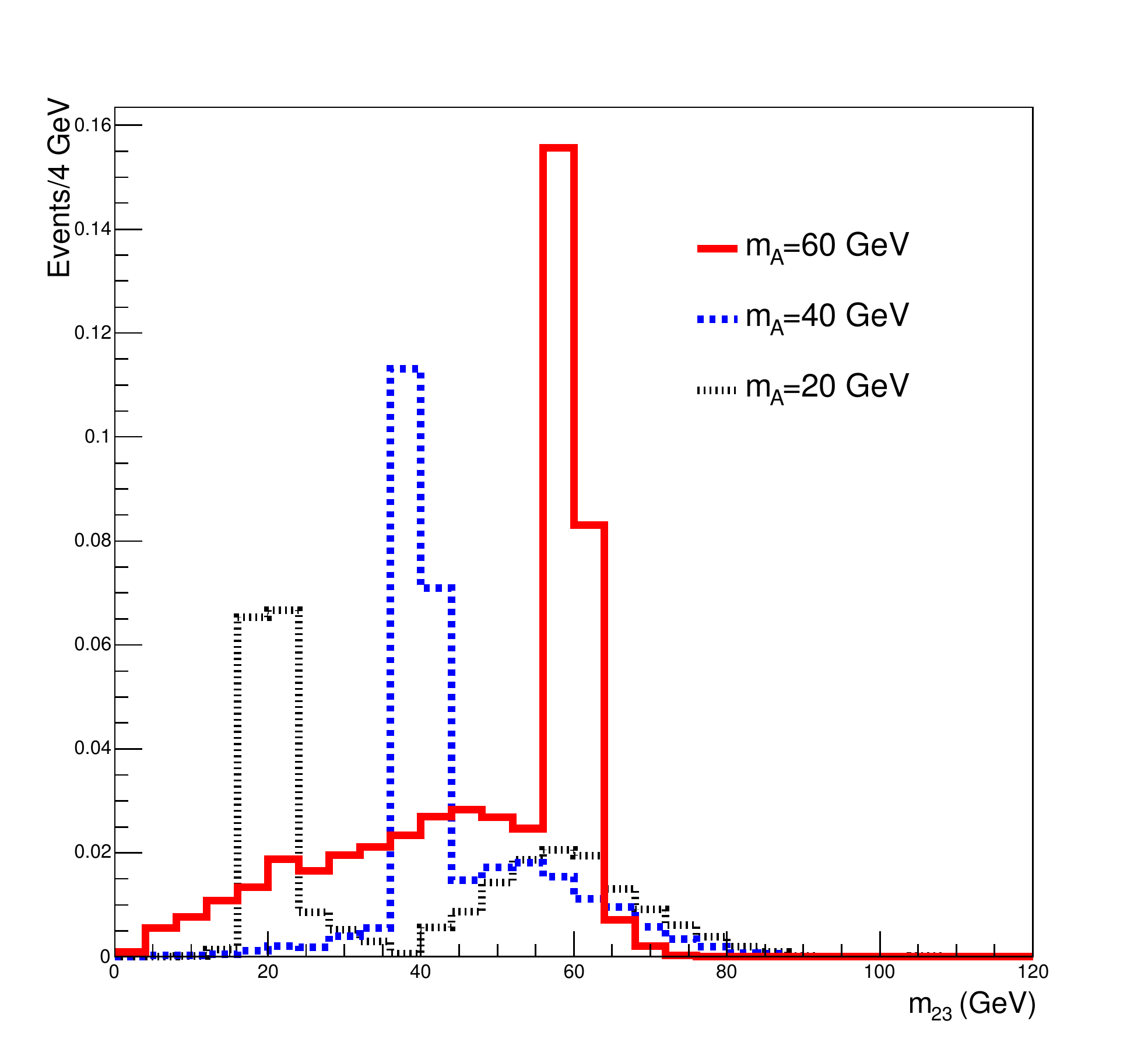}}
  \caption{Distributions at detector level:
    (a) $m_{3\gamma}$ for $gg\to H\to hh \to 4\gamma$, (b) $m_{3\gamma}$ for $gg\to H\to AA \to 4\gamma$, 
    (c) $m_{23}$ for $gg\to H\to hh \to 4\gamma$  and (d) $m_{23}$ for $gg\to H\to AA \to 4\gamma$.}\label{kin1}.
\end{figure}
In Fig. \ref{kin1}, we expose the similarity between the two processes by
showing some kinematic spectra of $gg \to H \to h h \to 4 \gamma$ and $gg \to
H \to A A \to 4 \gamma$. In particular, we present  the 
$m_{3 \gamma}$ spectrum (the invariant mass of the three leading
$P_t$-ordered photons) as well as  the $m_{23}$ spectrum (the invariant mass
of the 2nd and 3rd $P_t$-ordered photons). Obviously, these spectra show  no
significant difference for these two processes $gg \to H \to h h \to 4 \gamma$
and $gg \to H \to A A \to 4 \gamma$,  except fluctuations from numerical
simulation. 
Therefore, the experimental methods and results of multi-photon data from $gg
\to H \to AA \to 4 \gamma$ can also be applied to  $gg \to H \to h h \to 4
\gamma$.



In order to establish LHC sensitivity to our signal process, we determine the scaling factors for both signal and backgrounds, necessary to map our own MC simulations onto the real data results of ATLAS. In doing so, we use the Leading Order (LO) cross section to determine such factors. Therefore the latter should encode the $K$-factors due to 
higher order corrections, the difference between real detector and the fast detector simulations, the mistaging rate of a jet as a photon and an electron as a photon as well, etc.   In fact, when the rejection rate of a jet as a photon is considered  \cite{atlas1}, the fake rate could be around $10^{-3}$, so for the process $\gamma \gamma +j$ the scaling factor is dominantly determined by the fake rate while for the process $\gamma + jj$ we expect the fake rate to be around $10^{-6}$, the scaling factors demonstrating then that significant background contribution is indeed due to fake rates.

\begin{figure}[htbp]
  \centering
  \subfigure{
  \label{Fig3.sub.1}\thesubfigure
  \includegraphics[width=0.44\textwidth]{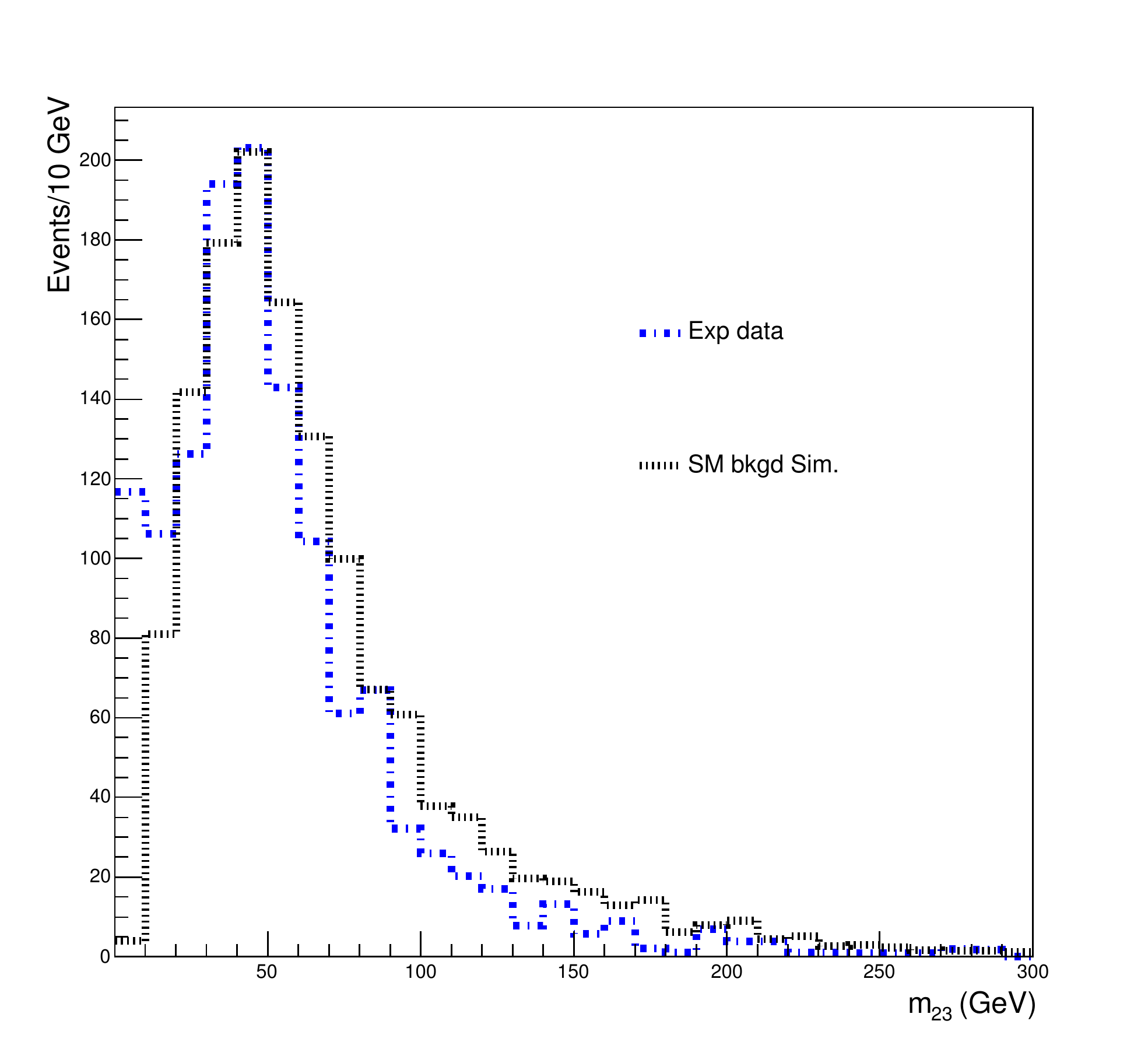}}
  \subfigure{
  \label{Fig3.sub.2}\thesubfigure
  \includegraphics[width=0.44\textwidth]{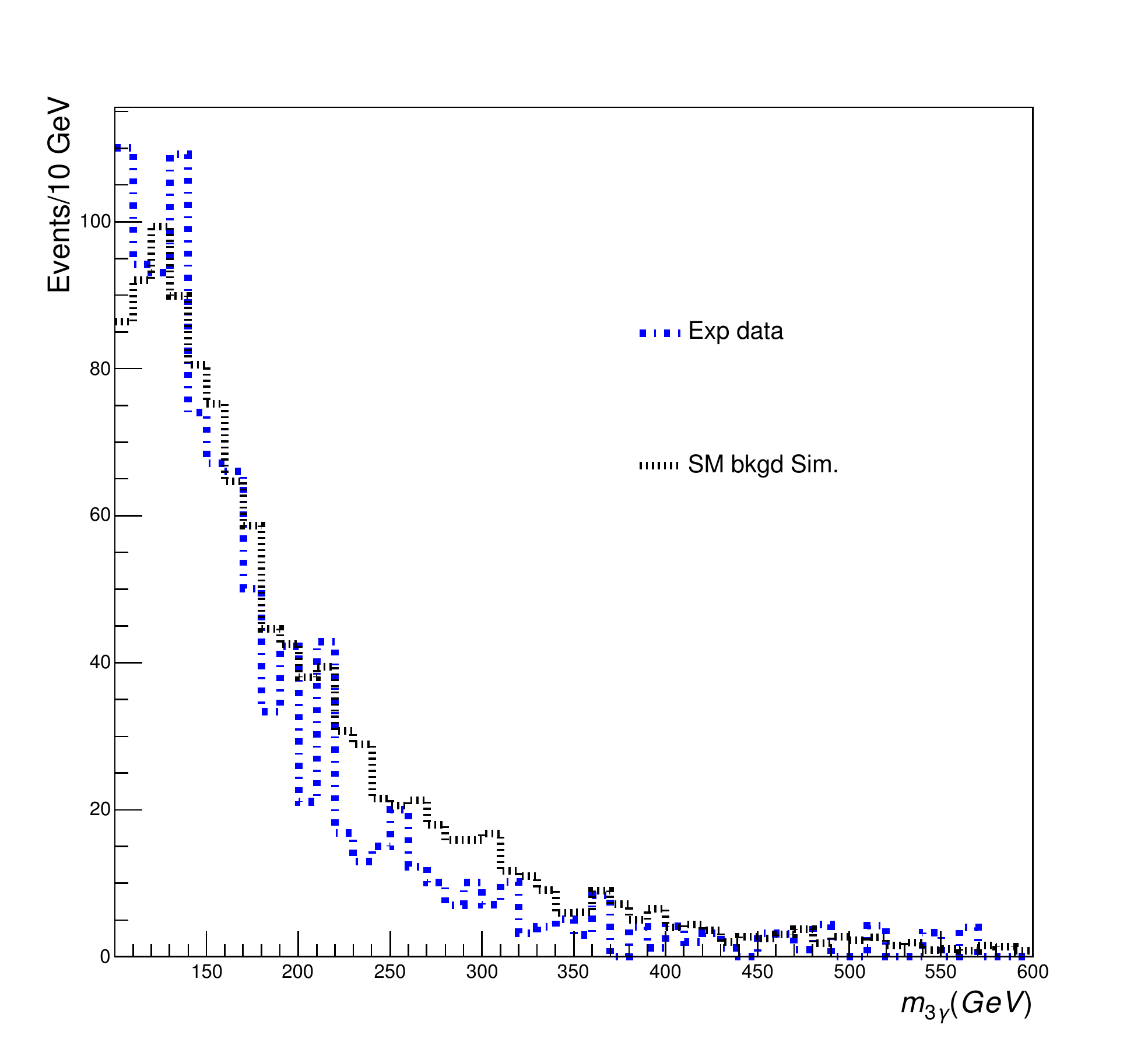}}
  \caption{The comparison of the simulated spectra of (a) $m_{23}$ and (b) $m_{3 \gamma}$ to those experimental ones is demonstrated.}\label{simexp}
\end{figure}

The scaling factors for each process are listed in Tab. \ref{scalingfactor}, as mentioned, being all determined from the aforementioned ATLAS results at 8 TeV. We also compare the experimental line-shapes with those from our MC events, which are plotted in Fig. \ref{simexp}. Although the spectra from MC are slightly harder and noticeable  differences appear  in the bins with $m_{3\gamma}<50$ GeV and $m_{2\gamma}<50$ GeV, the total number of predicted events is close to the experimental ones. 

By assuming the same scaling factors, we examine the boost factor in the LHC sensitivity for an increased collision energy of $\sqrt{s}=14$ TeV. The ensuing cross sections of the signal and  background processes are given in Tab. \ref{sen14}. Since the signal production process $gg\to H$ has a larger boost factor when the collision energy increases from 8 TeV to 14 TeV due to the larger enhancement of gluon flux, as compared to the more varied background composition,  it is natural to expect a better sensitivity for the future LHC runs, as readily seen in the table. For example, when the integrated luminosity of the  LHC  is assumed to be 300/fb, the boost factor in cross section (which is defined in the caption) is found to be 32.2 for thew signal and 25.7 for the background. This effect reflects then in  the projected sensitivities shown in Fig. \ref{fig4} for the LHC  with $\sqrt{s}=14$ TeV and 300/fb of luminosity (blue lines). 

\begin{center}
	\begin{table}
		\begin{center}
			\begin{tabular}{|c|c|c|c|c|c|c|}
				\hline
				process & $\sigma$ with $\sqrt{s}= 8$ TeV  &  N.o.E (Theory)    &  Acc. Eff.   & N.o.E (Expected)  &  N.o.E (Experimental) & Scaling factor   \\ 
				\hline
				$3\gamma$ & $72.5$ fb  & $1.47 \times 10^3$ & $18\%$ & $2.65 \times 10^2$ & $340\pm110$ & $1.28$  \\
				\hline 
				$2\gamma$ & $109$ pb  & $2.21 \times 10^6$ & $0.4\%$ & $8.8 \times 10^3$ & $330\pm50$ & $3.7 \times 10^{-2}$  \\
				\hline
				$2\gamma+j$ & $58.3$ pb  & $1.18 \times 10^6$ & $19\%$ & $2.24 \times 10^5$ & $350\pm50$ & $1.56 \times 10^{-3}$  \\
				\hline
				$\gamma+2j$ & $4.39 \times 10^4$ pb  & $8.91 \times 10^8$ & $15\%$ & $1.33 \times 10^8$ & $110\pm40$ & $8.31 \times 10^{-7}$  \\
				\hline
				$\gamma e^+e^-$ & $5.91$ pb  & $1.2 \times 10^5$ & $23.5\%$ & $2.8 \times 10^4$ & $89\pm11$ & $3.2 \times 10^{-3}$  \\
				\hline
				$2\gamma e^+e^-$ & $30$ fb  & $6.1 \times 10^2$ & $34\%$ & $2.07 \times 10^2$ & $85\pm22$ & $0.41$  \\
				\hline
				$\gamma W +X$ & $24.4$ pb  & $4.95 \times 10^5$ & $2.9\%$ & $1.43 \times 10^4$ & $11.4\pm1.5$ & $0.8 \times 10^{-3}$  \\
				\hline
			\end{tabular}
		\end{center}
		\caption{\label{scalingfactor} The scaling factors for each process of SM background are shown where N.o.E denotes the ``Number of Events" at 8 TeV  when the integrated luminosity is taken as  20.3/fb and where Acc. Eff. denotes 
the ``Acceptance Efficiency'' which 
 is determined by the selection cuts of the mentioned ATLAS analysis.} 
	\end{table}
\end{center}

\begin{center}
	\begin{table}
		\begin{center}
			\begin{tabular}{|c|c|c|c|}
				\hline
				& $\sigma$ with $\sqrt{s}= 8$ TeV  & $\sigma$ with $\sqrt{s}= 14$ TeV   &  Boost factor \\ 
				\hline
				Signal & 19.3 $\times \beta$ pb  & 42.0 $\times \beta$ pb &   32.2  \\
				\hline 
				Background & 67.5 fb  & 117 fb & 25.7   \\
				\hline
			\end{tabular}
		\end{center}
		\caption{\label{sen14} The projected LHC sensitivity at 14 TeV  is given, where $\beta={\rm BR}(H\to hh ) {\rm BR}^2(h\to \gamma \gamma)$, expressed in terms of the boost factor, defined as $\frac{\sigma(\sqrt{s}=\textrm{14 TeV}) \times L_{\textrm{14 TeV}}}{\sigma(\sqrt{s}=\textrm{8 TeV}) \times L_{\textrm{8 TeV}}}$, where $L_{\textrm{14 TeV}}$ is assumed to be 300/fb and $L_{\textrm{8 TeV}}$ is taken as 20.3/fb. } 
	\end{table}
\end{center}

\section*{Conclusions}
In this paper, we built upon previous results of some of ours, which had extracted a region of parameter space of the 2HDM-I where very light $h$ and $A$ states, down to 15--20 GeV or so, can be found, when the $H$ one is assumed to be the SM-like one discovered at the LHC in 2012. This spectrum is well compatible with all standard theoretical constraints (unitarity, vacuum stability, etc.) and all available experimental data (including flavour as well  as Higgs data) and thus offers the possibility of testing Higgs cascade decays of the type $H\to hh$ and $H\to AA$, compatibly with the total $H$ width extracted by global fits to the 125 GeV Higgs data. Amongst the possible decays of the $h$ and $A$ states we concentrated here upon those yielding di-photons, the overall signature then being a $4\gamma$ one, primarily induced by $gg\to H$ creation. We do so as the 2HDM-I can develop, over the aforementioned region of parameter space, a (nearly) fermiophobic limit, so that $h$ and $A$ decays into fermions (chiefly, $b\bar b$ and $\tau^+\tau^-$) are negligible. In fact, the availability of an ATLAS analysis performed on Run 1 samples of the LHC looking for these specific multi-photon signals allowed us, on the one hand, to validate our MC tools against the full detector environment of a multi-purpose LHC experiment and, on the other hand, to project our finding into the future by extrapolating our results to a collider energy of 14 TeV and luminosity of 300/fb. This exercise revealed that the portion of 2HDM-I parameter space where the above phenomenology is realised, while being just below the current LHC sensitivity, is readily accessible at future stages of the LHC. To confirm or disprove its existence is of paramount importance as this would almost univocally point to a specific realisation of a generic 2HDM construct as such light and fermiophobic $h$ and $A$ states cannot be realised in alternative formulations of it.  

\section*{Acknowledgements}
AA, RB and SM  are supported by the grant H2020-MSCA-RISE-2014 no. 645722
(NonMinimalHiggs). This work is also supported by the Moroccan Ministry of Higher
Education and Scientific Research MESRSFC and  CNRST: Projet PPR/2015/6.
SM is supported in part through the NExT Institute. 
RB is supported in part by the Chinese Academy of Sciences (CAS) President's International Fellowship Initiative (PIFI) program (Grant No. 2017VMB0021). Q.S. Yan and X.H. Zhang are supported by the Natural Science Foundation of China
under the grant no. 11575005.

\appendix
\section{The squared matrix elements of $gg \to H \to hh \to 4 \gamma$  and $gg \to H \to AA \to 4 \gamma$ }
\label{appenda}

In this appendix, we demonstrate in details that the process $gg \to H \to hh \to 4 \gamma$ and the process $gg \to H \to A A \to 4 \gamma$ have the same differential cross section, except an overall factors.

For the CP-even Higgs case, the matrix element of the process $gg \to H \to hh \to 4 \gamma$ can be put as
\begin{align}
	\nonumber i \mathcal{M} &= iC(k_1\cdot k_2\eta^{\mu\nu}-k^\mu_2k^\nu_1)\epsilon^*_\mu(k_1)\epsilon^*_\nu(k_2)(k_3\cdot k_4\eta^{\rho\sigma}-k^\rho_4k^\sigma_3)\\
	  &\quad \times\epsilon^*_\rho(k_3)\epsilon^*_\sigma(k_4)\delta^{ab}
	\epsilon(p_1)\cdot \epsilon(p_2)\,,
\end{align}
where $p_1$ and $p_2$ is the momentum of the initial gluons, $k_1-k_4$ are momentum of 4 photons in the final state. Note that we have group the  effective couplings of each vertices and the propagators together in $C$ and just show their Lorentz structure and color indices. The squared amplitude with the average of the degree of freedom of initial states and the sum over the polarisations of photons in the final states yields
  \begin{align}
	\nonumber \overline{|\mathcal{M}|^2} &= \frac{1}{2}\frac{1}{2}\frac{1}{8}\frac{1}{8}\sum_{pols}\sum_{a,b=1}^8|\mathcal{M}|^2 \\ \nonumber &=  \frac{1}{256}|C|^2\delta^{ab}\delta_{ab}(k_1\cdot k_2\eta^{\mu\nu}-k^\mu_2k^\nu_1)
	(k_1\cdot k_2\eta^{\mu'\nu'}-k^{\mu'}_2k^{\nu'}_1)
	\\ \nonumber &\quad \times \sum_{\epsilon}\epsilon^*_\mu(k_1)\epsilon_{\mu'}(k_1) \sum_{\epsilon}\epsilon^*_\nu(k_2)\epsilon_{\nu'}(k_2)(k_3\cdot k_4\eta^{\rho\sigma}-k^\rho_4k^\sigma_3)\\\nonumber &\quad \times(k_3\cdot k_4\eta^{\rho'\sigma'}-k^{\rho'}_4k^{\sigma'}_3)
	 \sum_{\epsilon}\epsilon^*_\rho(k_3)\epsilon_{\rho'}(k_3)\sum_{\epsilon}\epsilon^*_\sigma(k_4)\epsilon_{\sigma'}(k_4)\\ &\quad \times
	\sum_{\epsilon}|\epsilon(p_1)\cdot \epsilon(p_2)|^2\,.
  \end{align}
	By using the polarisation sum formula
	\begin{eqnarray}
		\sum_{\epsilon}\epsilon^*_\mu\epsilon_\nu=-\eta_{\mu\nu}\,,
	\end{eqnarray} 
and the fact $\delta_{ab}\delta^{ab}=8$, we arrive at  
	\begin{eqnarray}
		\overline{|\mathcal{M}|^2}=\frac{1}{16}|C|^2[2(k_1\cdot k_2)^2-k_1^2k_2^2][2(k_3\cdot k_4)^2-k_3^2k_4^2]\,.
	\end{eqnarray}
	Due to the on-shell conditions for the photons of the final state, we have  $k_1^2=k_2^2=k_3^2=k_4^2=0$, then we obtain the following squared matrix element 
	\begin{eqnarray}
	\overline{|\mathcal{M}|^2}=\frac{1}{4}|C|^2(k_1\cdot k_2)^2(k_3\cdot k_4)^2 \label{hhcase}\,.
	\end{eqnarray}

For the CP-odd Higgs case, the matrix element of the process $gg \to H \to AA \to 4 \gamma$ can be put as
	\begin{eqnarray}
		i \mathcal{M} =iD\epsilon^*_\alpha(k_1)\epsilon^*_\beta(k_2)\epsilon^{\alpha\beta\mu\nu}k^1_\mu k^2_\nu\epsilon^*_\rho(k_3)\epsilon^*_\sigma(k_4)\epsilon^{\rho\sigma\gamma\delta}k^3_\gamma k^4_\delta\delta^{ab}
		\epsilon(p_1)\cdot \epsilon(p_2)\,,
	\end{eqnarray}
where D includes all effective couplings and the propagators. The squared matrix element with a sum over the polarisations of photons in the final state is given as 
	\begin{align}
		\overline{|\mathcal{M}|^2} &= \frac{1}{2}\frac{1}{2}\frac{1}{8}\frac{1}{8}\sum_{pols}\sum_{a,b=1}^8|\mathcal{M}|^2  \nonumber  
		\\ &= \frac{1}{256}|D|^2\delta_{ab}\delta^{ab}\sum_{\epsilon}\epsilon^*_\alpha(k_1)\epsilon_{\alpha'}(k_1)\sum_{\epsilon}\epsilon^*_\beta(k_2)\epsilon_{\beta'}(k_2)\epsilon^{\alpha\beta\mu\nu}k^1_\mu k^2_\nu \nonumber 
		\\ &\quad \times \nonumber\epsilon^{\alpha'\beta'\mu'\nu'}  k^1_{\mu'} k^2_{\nu'}\sum_{\epsilon}\epsilon^*_\rho(k_1)\epsilon_{\rho'}(k_1)\sum_{\epsilon}\epsilon^*_\sigma(k_2)\epsilon_{\sigma'}(k_2)\epsilon^{\rho\sigma\gamma\delta}k^1_\gamma k^2_\delta \nonumber 
		\\ &\quad \times \epsilon^{\rho'\sigma'\gamma'\delta'}k^1_{\gamma'} k^2_{\delta'} \sum_{\epsilon}|\epsilon(p_1)\cdot \epsilon(p_2)|^2
	    \nonumber \\ &= \frac{1}{32}|D|^2\eta_{\alpha\alpha'}\eta_{\beta\beta'}\epsilon^{\alpha\beta\mu\nu}k^1_\mu k^2_\nu\epsilon^{\alpha'\beta'\mu'\nu'}  k^1_{\mu'} k^2_{\nu'} \nonumber 
	    \\ &\quad \times \eta_{\rho\rho'}\eta_{\sigma\sigma'}\epsilon^{\rho\sigma\gamma\delta}k^1_\gamma k^2_\delta\epsilon^{\rho'\sigma'\gamma'\delta'}k^1_{\gamma'} k^2_{\delta'} \nonumber 
	    \\ &\quad \times  (|\epsilon^{(+)}(p_1)\cdot \epsilon^{(-)}(p_2)|^2+|\epsilon^{(-)}(p_1)\cdot \epsilon^{(+)}(p_2)|^2) \nonumber 
	    \\ &= \frac{1}{16}|D|^2\epsilon^{\alpha\beta\mu\nu}k^1_\mu k^2_\nu\epsilon^{\alpha\beta\mu'\nu'}  k^1_{\mu'} k^2_{\nu'}  \epsilon^{\rho\sigma\gamma\delta}k^1_\gamma k^2_\delta\epsilon^{\rho\sigma\gamma'\delta'}k^1_{\gamma'} k^2_{\delta'} \nonumber 
	    \\ &= \frac{1}{16}|D|^22[(k_1\cdot k_2)^2-k_1^2 k_2^2]2[(k_3\cdot k_4)^2-k_3^2k_4^2] \nonumber
	    \\ &= \frac{1}{4}|D|^2(k_1\cdot k_2)^2(k_3\cdot k_4)^2 \label{aacase}\,.
	\end{align}
Since both squared amplitudes given in Eq.(\ref{hhcase}) and Eq. (\ref{aacase}) are proportional to the factor $(k_1\cdot k_2)^2(k_3\cdot k_4)^2$, while the rest of factors could not lead to a significantly different kinematic dependence (except the propagators are different for these two processes where the total decay widths of h and A are different.), therefore there is no surprise to observe that the kinematics of these two processes are the same when the polarisations of photons are summed over.

 \end{document}